%% file: main.tex
\documentclass[sigconf,urlbreakonhyphens,pbalance]{acmart}

\newif\ifanonymize
\anonymizefalse  %

\copyrightyear{2025} 
\acmYear{2025} 
\setcopyright{cc}
\setcctype{by}
\acmConference[ASIA CCS '25]{ACM Asia Conference on Computer and Communications Security}{August 25--29, 2025}{Hanoi, Vietnam}
\acmBooktitle{ACM Asia Conference on Computer and Communications Security (ASIA CCS '25), August 25--29, 2025, Hanoi, Vietnam}\acmDOI{10.1145/3708821.3733890}
\acmISBN{979-8-4007-1410-8/2025/08}

\usepackage{subcaption}
\usepackage{calc}
\usepackage{amstext}
\usepackage{amsmath}
\usepackage{amsthm}
\usepackage{float}
\usepackage{mathtools}

\newcommand{\todo}[1]{}

\newcommand{\todoi}[1]{\todo[inline]{#1}}

\newtheoremstyle{mystyle}%
  {\topsep}%
  {\topsep}%
  {\itshape}%
  {}%
  {\scshape}%
  {.}%
  {.75em}%
  {}%
\theoremstyle{mystyle}
\newtheorem{lemma}{Lemma}

\usepackage[utf8]{inputenc}
\newcommand{\enquote}[1]{``#1''}
\usepackage{url}

\usepackage{siunitx}

\usepackage[inline]{enumitem}

\usepackage{listings}
\usepackage{makecell}
\usepackage{xspace}
\usepackage{float}
\usepackage{textcomp}

\usepackage{graphicx}

\usepackage{algorithm}%
\usepackage{algpseudocode}%
\algblock{Input}{EndInput}
\algnotext{EndInput}
\algblock{Output}{EndOutput}
\algnotext{EndOutput}

\usepackage{stfloats}
\usepackage{newfloat}

\usepackage{tikz}

\usepackage{microtype}

\usepackage{hyperref}
\usepackage{cleveref}

\DeclareFloatingEnvironment[name=Protocol]{protocol}
\crefname{protocol}{Protocol}{Protocols}
\DeclareCaptionFormat{algor}{%
  \hrulefill\par\offinterlineskip\vskip1pt%
    \textbf{#1#2}#3\offinterlineskip\hrulefill}
\DeclareCaptionStyle{algori}{singlelinecheck=off,format=algor,labelsep=space}
\newcommand*{\priority}[1]{\protect\tikz[baseline=-3pt, scale=0.11]{%
        \protect\draw (0,0) circle (1);%
        \fill[fill opacity=0.5,fill=black] (0,0) -- (90:1) arc (90:90-#1*3.6:1) -- cycle;%
    }%
  }
\newcommand{\fully}{\priority{100}}
\newcommand{\partially}{\priority{50}}
\newcommand{\unsupported}{\priority{0}}

\newcommand{\rom}[1]{(\uppercase\expandafter{\romannumeral #1\relax})}

\newcommand{\AUDSTR}{\mathsf{audience}}
\newcommand{\AUDHASH}{\mathsf{AudienceId}}

\newcommand{\OPRF}{\mathsf{OPRF}}
\newcommand{\BLIND}{\mathsf{Blind}}
\newcommand{\EVAL}{\mathsf{BlindEval}}
\newcommand{\UNBLIND}{\mathsf{Unblind}}

\newcommand{\UID}{\mathsf{userId}}

\newcommand{\R}{r}
\newcommand{\BLINDINPUT}{A}
\newcommand{\BLINDOUTPUT}{B}

\newcommand{\POINTSGROUP}{\mathbb{G}}
\newcommand{\POINTS}{\POINTSGROUP^+}
\newcommand{\ORDER}{p}
\newcommand{\SCALARS}{\mathbb{Z}_\ORDER^\times}

\newcommand{\myparagraph}[1]{\textbf{#1.\hspace{.5em}}}

\ifanonymize
  \newcommand{\THEREPO}{\url{https://anonymous.4open.science/r/AsiaCCS25BISON} \emph{(blinded)}}
\else
  \newcommand{\THEREPO}{\url{https://github.com/iaik-jheher/BISON}}
\fi

\begin{document}

\date{\today}
\title[BISON: Blind Identification with Stateless scOped pseudoNyms]{BISON: Blind Identification\\ with Stateless scOped pseudoNyms}

\author{Jakob Heher}
\orcid{0009-0007-7351-8117}
\affiliation{%
  \institution{A-SIT, Graz University of Technology}
  \city{Graz}
  \country{Austria}
}

\author{Stefan More}
\orcid{0000-0001-7076-7563}
\affiliation{%
  \institution{A-SIT, Graz University of Technology}
  \city{Graz}
  \country{Austria}
}

\author{Lena Heimberger}
\orcid{0009-0001-9404-7699}
\affiliation{%
  \institution{Graz University of Technology}
  \city{Graz}
  \country{Austria}
}

\begin{abstract}
  Delegating authentication to identity providers like Google or Facebook, while convenient, compromises user privacy.
  These identity providers can record users' every move; the global identifiers they provide also enable internet-wide tracking.

  We show that neither is a necessary evil by presenting the BISON pseudonym derivation protocol, inspired by Oblivious Pseudorandom Functions.
  It hides the service provider's identity from the identity provider yet produces a trusted, scoped, immutable pseudonym.
  Colluding service providers cannot link BISON pseudonyms; this prevents user tracking.
  BISON does not require a long-lived state on the user device and does not add additional actors to the authentication process.

  BISON is practical.
  It is easy to understand, implement, and reason about, and is designed to integrate into existing authentication protocols.
  To demonstrate this, we provide an OpenID Connect extension that allows OIDC's PPID pseudonyms to be derived using BISON while remaining fully backwards compatible.
  Additionally, BISON uses only lightweight cryptography.
  Pseudonym derivation requires a total of four elliptic curve scalar-point multiplications and four hash function evaluations, taking $\approx$3 ms in our proof of concept implementation.
  Thus, BISON's privacy guarantees can be realized in practice.
  This makes BISON a crucial stepping stone towards the privacy-preserving internet of tomorrow.

\end{abstract}

\maketitle

\section{Introduction}
\label{0:introduction}

In today's internet, nary a day passes without being prompted to create an account for a newsletter, a website, or simply to keep reading an interesting article.
Users tend to value comfort and usability, and just want to get back to what they were doing.
They will often take the more convenient route and just \enquote{Sign in with Google}, \enquote{Log in with Facebook}, or \enquote{Sign in with Apple} \cite{DBLP:conf/uss/GhasemisharifRC18,DBLP:journals/corr/abs-2302-01024}.
The promise is as simple as it is tempting:
  forget remembering all these different passwords -- access the entire internet, with just one set of credentials to worry about.

Of course, the flip side isn't talked up as much.
The service serving as \enquote{your identity} gets involved every time you log in -- anywhere.
Highly personalized association data is at its fingertips, including valuable information about users' behavior;
  attractive for profiling, marketing, and worse.
In the internet of authentication, it's no surprise that everyone wants to be the identity provider.

But, do we need this?
Is it really necessary for identity providers to learn %
which services their users frequent?
Can privacy only be provided by highly complex systems struggling to find real-world adoption?
In this work, we can answer these questions with a resounding \enquote{no}.
It does not have to be this way.

\subsection{Challenges}

Delegated authentication faces two fundamental challenges:

\myparagraph{Challenge 1: Portability}
Users are fickle, unpredictable beings~\cite{abel2012consumer}.
They don't just use their accounts on a single, stable device, or even on some predictable set of user devices.
Sometimes they'll log in from their friends' house, or a hotel lobby; from a shared computer at the school library, or simply a private browsing session.

In all of these cases, they still expect their accounts to be there, just one short authentication process away~\cite{DBLP:conf/soups/SunPMDHB11,gafni2014social}.
This reality is in sharp contrast with some academic work~\cite{DBLP:conf/ccs/CamenischH02,DBLP:conf/securecomm/MostowskiV11,DBLP:journals/popets/ZhangKSZR21,DBLP:conf/compsac/PodgorelecAZ22,DBLP:conf/soups/KorirPD22}, which envisions secret key material being stored \enquote{on the user's device}.
Without this material, authentication is impossible.

In practice, this is a problem: a seamless user experience is a baseline requirement.
In this work, we will thus instead aim for a reality-proof solution, by embracing \emph{statelessness} of the user device.

\myparagraph{Challenge 2: Privacy}
\emph{Who you talk to} is highly sensitive.
Maybe you're getting confidential medical advice from a specialist, seeking aid with interpersonal conflict, or reaching out to a self-help group.
Either way, nobody except \emph{you} and \emph{them} should know about it;
  the two of you should enjoy \emph{relationship anonymity}~\cite{anonTerminology,DBLP:reference/crypt/Bleumer11af,DBLP:reference/crypt/X11ko}.

At the same time, \emph{who you are} is also highly sensitive information.
Let's say you log in to an internet forum, and you log in to a shopping platform.
The shopping platform might be interested in the messages you post on the forum, so they can know which products to advertise to you~\cite{DBLP:journals/mansci/KramerSW19}.
Maybe you even want this, and might consent to it.
But these parties should not be able to correlate your information automatically;
  you should enjoy \emph{pseudonym unlinkability}~\cite{DBLP:conf/uss/RaoR00,DBLP:conf/pet/SteinbrecherK03,anonTerminology,DBLP:reference/crypt/Bleumer11ae}.
In this work, we will aim to safeguard \emph{both} of these privacy properties.
As we will outline in more detail in \Cref{1:goals}, this is an especially tricky challenge;
  many na\"ive solutions to one problem simultaneously preclude solving the other.

\subsection{Contributions}
We introduce BISON, a stateless pseudonym derivation protocol that prevents profiling by identity providers and tracking by service providers.
It produces pseudonyms that are scoped, immutable, and stable across multiple authentications.

BISON takes inspiration from Oblivious Pseudo-Random Functions (OPRFs), a well-researched cryptographic primitive, and applies it to a novel context.
As a result, it is straightforward to understand and implement.
Additionally, BISON is lightweight, only requiring four hash function evaluations and four elliptic curve point-scalar multiplications per derivation.
It does not introduce any additional actors into the established authentication flow, does not require any long-lived state on the user device, and can operate on well-supported groups such as NIST P-256.
We provide a security proof, showing that BISON exhibits the claimed properties.

We then show that the existing OpenID Connect protocol, widely used in federated authentication, can be easily augmented to use BISON for pseudonym derivation.
This allows user agents to take an active role in safeguarding user privacy in federated authentication.
As browser vendors are currently deliberating the adoption of novel solutions to federated authentication across isolated browsing contexts, this further demonstrates that BISON can be practically deployed in real-world scenarios.
Our extension is backwards-compatible, falling back to unaugmented OpenID Connect.
We provide a full demonstrator, and make its source code available.

\subsection{Outline}
The remainder of this work is composed of two main parts.
First, in \Cref{1:section}, we describe \textbf{the abstract BISON protocol};
  we introduce our goals (\Cref{1:definitions,1:goals,1:threat-model}) and necessary background (\Cref{1:background:oprf}), then present the BISON protocol (\Cref{1:protocol}).
  In \Cref{sec:evaluation}, we \textbf{prove BISON's security} (\Cref{1:analysis}) and discuss external considerations (\Cref{1:analysis:external}) and protocol details (\Cref{1:discussion}).
Next, in \Cref{2:section}, we \textbf{integrate BISON derivation into OpenID Connect};
  we introduce OpenID Connect (\Cref{2:background}), specify our extension (\Cref{2:oidc-extension}) and how it meets security requirements (\Cref{2:analysis}).
  We describe our implementation (\Cref{2:implementation}), show that it has negligible overhead (\Cref{2:evaluation}), and discuss the implications (\Cref{2:discussion}).
Finally, we compare to existing solutions in \Cref{related-work}. %

\begin{figure}[H]
    \centering
    \includegraphics[width=\columnwidth]{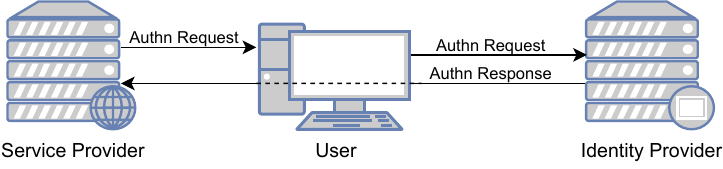}
    \caption{The basic authentication flow. Using BISON derivation does not alter it; we do not add actors or messages.}
    \label{fig:arch}
\end{figure}
\section{The BISON Protocol}
\label{1:section}
In this first part of our work, we introduce the BISON protocol for pseudonym derivation.
It offers \emph{pseudonym unlinkability} against colluding service providers, \emph{User-SP relationship anonymity} against a malicious identity provider, and \emph{Sybil resistance} against a malicious user, while requiring \emph{no persistent state} on the user device.

\subsection{Definitions}
\label{1:definitions}

We first define the three protocol actors and the roles they fulfill, and make some clarifications regarding terminology and notation.
Note that we do not require actors beyond those common in existing authentication protocols~\cite{OIDC, Shibboleth}.

\myparagraph{User/User Device}
The \emph{user} is a party with the primary objective of proving their identity to some service securely and efficiently, ideally in a privacy-preserving manner.

When necessary, we will distinguish between the human \enquote{user}, and the \enquote{user device} they are operating, which will, for example,, perform cryptographic operations on their behalf.

\myparagraph{Service Provider (SP)}
The \emph{service provider} is a party that requires the user to authenticate, as part of an authorization decision to access some service or resource.
There are many different SPs, which may offer wildly disparate services.
An internet message board, a virtual storefront, a health provider's appointments software, an online banking application, and a government bureau's web portal, are very different kinds of SPs, with different needs.

Almost all SPs share a need to (re-)identify users.
If a user visits a message board, they should have the ability to edit messages they have previously sent;
if they visit a storefront, they should be able to see their pending orders;
if they visit their health provider, they should be able to see and cancel their appointments;
and so forth.
It is thus necessary to perform an authentication process of some kind, which results in a trusted persistent identifier for the user.

Authenticating a user is not a trivial task.
Depending on the method, the server might need to securely store persistent information -- passwords, shared secrets, or recovery codes -- for each user.
Conversely, users will then also need to remember many different credentials for different service providers.

To summarize, authentication is a burdensome task for both service providers and users.
This has led to a desire to delegate the actual authentication process to a dedicated third party.

\myparagraph{Identity Provider (IdP)}
The \emph{identity provider} is a party that is trusted to perform user authentication.
It decides on the authentication factors to use, and securely stores any information necessary to \mbox{(re-)}authenticate the user using these factors, such as password digests, shared secrets, or public keys.

The identity provider then assigns each \emph{user account} a \enquote{global} identifier, which is only unique within the context of this particular IdP.
Some identity providers may use truly global identifiers, such as email addresses;
  but not all do.
Depending on the context, possession of an account at a certain identity provider may carry particular implications, such as membership in a certain organization, which is verified by the identity provider.
In such a case, the service provider delegates not only authentication, but implicitly also verification of these properties, to the identity provider.

\myparagraph{Audience}
In BISON, users hold a particular stable, unique, and non-reassignable pseudonym which is scoped to a specific \emph{audience}.

What constitutes an audience scope is highly context-specific.
At a na\"ive level, each SP might form its own audience.
However, related SPs, which should be permitted to share data, might be grouped to form a single audience.
By contrast, a single SP might also be split into different audiences, for example to prevent data aggregation across divisions of a large corporation.

A particular audience is identified by a byte string, called an \emph{audience identifier};
  going forward, we will refer to these identifiers as $\AUDSTR$s for brevity.
In BISON, we assume that the user device can use some context-specific method to determine whether a service provider is authorized for a particular $\AUDSTR$ value.
For example, this decision might be based on a certificate issued by the identity provider, or it might be based on already-available trusted information.
We discuss this requirement further in \Cref{1:analysis:sp-authentication}.

\myparagraph{Alternate Terminology}
We have chosen the terms above for use throughout this work in an attempt to avoid ambiguity.
However, identity management is a broad field, and different sub-fields have adopted different terminologies for related concepts.
To ease the reader's understanding, we briefly mention such alternative terms that may appear in related work.

The \emph{user device} may also be referred to as a \emph{client}, a \emph{user agent}, or a \emph{wallet}. The \emph{user} may be called \emph{resource owner}.

The \emph{service provider} is often also called a \emph{relying party}.
In OpenID Connect-related contexts, the service provider may -- somewhat confusingly -- also be referred to as a \emph{client}~\cite{OIDC}.

The \emph{identity provider} is, based on context, referred to as \emph{(authorization) server}, \emph{(credential) issuer}, or \emph{(claims) provider}.

\myparagraph{A Note on Notation}
For the sake of clarity, we briefly outline the mathematical notation used throughout the remainder of this work.
$(\POINTSGROUP, +)$ is a group of prime order $\ORDER = \mathsf{ord}(\POINTSGROUP)$.
The group operation is written additively.
We use $\POINTS$ to refer to the non-identity elements of $\POINTSGROUP$, i.e., $\POINTS := \POINTSGROUP \setminus \{ {0}_\POINTSGROUP \}$.
The repeated application of the group operation, written as $s \cdot X$ for $s \in \SCALARS, X \in \POINTSGROUP$, corresponds to $X + X + X + X + \ldots$ repeated $s$ times.
We will use uppercase (such as $X$ or $\AUDHASH$) to denote elements of $\POINTS$, and lowercase (such as $\R$ or $\UID$) for elements of $\SCALARS$.

\myparagraph{Hardness Assumptions}
We assume that, on $\POINTSGROUP$, both the Decisional Diffie-Hellman Problem (DDHP) and the Discrete Logarithm Problem (DLP) are computationally infeasible for an attacker to decide/solve. This common assumption is widely believed to hold for cryptographic groups such as EC groups and $(\mathbb{Z}/n\mathbb{Z})^\times$~\cite{DDHP,DLP}.

\subsection{Design Goals}
\label{1:goals}
We now explicitly state our goals for the protocol.
In doing so, we will use the terminology of Pfitzmann and Hansen~\cite{anonTerminology}.

\myparagraph{Pseudonym Unlinkability}
Delegating authentication to a central provider is convenient, but enabling a link between otherwise-disparate user accounts is a significant privacy risk.

At a basic level, consider a \enquote{normal} centralized authentication process, as shown in \Cref{fig:arch}.
The SP redirects the user to the identity provider, and the user authenticates there.
The SP is then provided with some assurance, signed by the IdP, that the user was authenticated.
This includes the user's identifier, which is commonly a (unique) account name, email address, or similar.
The same user identifier is then provided to each SP that the user authenticates to.
This allows different SPs, or other entities that they share data with, to link together the user's actions without their consent.

Commonly, there is no legitimate need for such cross-provider tracking.
Thus, \emph{in general}, there is no reason for every SP to obtain this same global identifier.
Some authentication delegation schemes recognize this, and support SP-specific pseudonyms \cite{OIDC, Shibboleth}.
Here, the IdP generates a persistent pseudonymous identifier for each SP.
A given user's identifiers at different SPs are unrelated to one another, and cannot be linked.
Thus, they cannot be used for tracking.

We formalize this concept of \emph{pseudonym unlinkability} as follows:
Two SPs, $A$ and $B$, each request authentication, which is completed by user accounts $X$ and $Y$ respectively.
As a result, they obtain pseudonyms $P_{X,A}$ and $P_{Y,B}$.
It should be impossible for them to decide whether the two pseudonyms refer to the same user account ($X = Y$), or to different user accounts ($X \neq Y$).

\myparagraph{User-SP Relationship Anonymity}
The use of a central IdP also means that this IdP will commonly learn the identity of any service provider(s) that the user authenticates to.
This is, once again, a significant privacy risk, with IdPs gathering a significant amount of data regarding their users' associations.

An IdP necessarily needs to identify the particular user that is authenticating.
Furthermore, it is generally legitimate for the IdP to control which SPs can use its authentication service.
However, it is \emph{not} generally necessary for the IdP to learn \emph{which particular user} is authenticating to \emph{which particular SP}.
The sensitive information at risk is the link between the user and a particular SP and user.

This, too, is a privacy risk that is well-known, both in literature and practice;
ways to authenticate at an unknown SP already exist \cite{corella2011nstic, BrowserID, DBLP:conf/ccs/FettKS15}.
Yet, combining this property with the previous \emph{pseudonym unlinkability} goal offers a unique challenge.
After all, if the identity provider does not know the SP's identity, how can it derive a SP-specific pseudonym for the user?

We formalize this concept of \emph{relationship anonymity} as follows:
An authentication request $R$ is received by the IdP.
They should learn no additional information about the SP's identity from this;
  in terms of probabilities, for each service provider $A$, $\mathbb{P}(A | R) = \mathbb{P}(A)$.

\myparagraph{Sybil Resistance}
It is generally desirable for many internet services to place some barrier on the creation of fresh user accounts.
Failing to do so may lead to the proliferation of spam, subversion of reputation or accountability systems, and other related issues.
This is commonly referred to as a \emph{Sybil attack}, and systems preventing this are termed \emph{Sybil-resistant}~\cite{DBLP:conf/iptps/Douceur02}.

Services that try to obtain Sybil-resistance commonly require some in-person contact or a verifiable link to a real-world identifier that is not easy to obtain.
Examples include phone number verification, scans of identification documents, or in-person account registration.
By their very nature, these interactions are highly invasive to a user's privacy.

Some approaches to privacy-preserving authentication attempt to prevent user tracking by both the SP and the IdP.
They do this by relying on the user to associate pseudonyms to particular service providers.
In such a scheme, the IdP simply allows the user to authenticate using one of any number of pseudonyms;
the user device is then expected to re-use the same pseudonym when re-authenticating to the same SP~\cite{PseudoID10}.

This completely voids any Sybil-resistance efforts by the IdP.
Since a single user account can present multiple unlinkable identifiers towards the SP, a single verified user at the IdP can spawn an arbitrary number of \enquote{identities} at the SP.
Thus, the SP would need to begin its own Sybil-resistance efforts.
This defeats the point of delegated authentication;
  and, by necessitating an invasive verification process, nullifies any privacy gains towards the SP.

We thus aim for \emph{Sybil resistance}, formalized as follows:
Given user account $X$ and SP $A$, there exists some stable pseudonym $P_{X,A}$ such that an authentication by $X$ at $A$ always results in $P_{X,A}$.

\myparagraph{No persistent user device state}
While storing secret information on the user's device can be very convenient for protocol design, it presents many practical challenges.

Devices may break or be misplaced, leading to the irrevocable loss of any cryptographic information stored.
Backing up such information while keeping it truly private is an open challenge, to which no perfect solution appears to exist \cite{zinkus2021sok}.

Users may also wish to log in on a shared computer, or on a friend's device.
The process of transferring cryptographic key material to such a device is often complex, and exposing key material to a potentially-untrusted device is undesirable.

We side-step all of these issues by requiring no persistent state on the user device.
All state kept on the user device should be ephemeral to a single authentication process, and should be able to be discarded after the process completes.
This matches existing protocols deployed in the real world, and avoids complications when integrating our scheme.

\myparagraph{Soundness \& Validity}
Finally, we explicitly state the two implicit goals of any delegated authentication scheme.

\emph{Soundness}:
If a user successfully completes an authentication process, this should guarantee that they have successfully authenticated to the IdP as the user corresponding to the resulting pseudonym.
A malicious user should not be able to -- without the IdP's cooperation -- impersonate benign users towards a SP.

\emph{User account validity}:
It should be possible for the IdP to suspend or delete a user account, disabling that user's ability to authenticate.

\subsection{Threat Model}
\label{1:threat-model}

We assume the \emph{user} to be fully malicious.
We consider two possible goals for a malicious user.
First, obtaining a different pseudonym for the same user-audience combination.
  This can either be the pseudonym of a particular genuine user (targeted), or just a blank-slate Sybil identity (untargeted).
Second, completing the authentication without querying the identity provider, while resulting in the user's genuine pseudonym.
  This would then allow a user to complete authentication using, for instance, a suspended user account.

We assume the \emph{service provider} to be fully malicious.
We consider three possible goals of malicious service providers.
First, obtaining the user's global identifier, which may be identifying information such as an email address.
Second, correlating the pseudonyms of users across different $\AUDSTR$s.
Third, impersonating a user towards another service provider.

We do not consider the implications of the service provider and identity provider colluding against the user.
As we argue in \Cref{1:discussion:unmasking}, in authentication protocols without long-lived user device state, such collusion can always de-anonymize the user.

The \emph{identity provider} is tasked with performing opaque, context-specific steps to authenticate the user.
They are the ultimate authority for who a user is.
Them being able to issue false attestations, or create make-believe identities, is baked into the fundamental assumptions of the system.
Thus, a fully malicious identity provider can trivially violate many of the goals of an authentication scheme.

We will therefore assume that the identity provider is honest(-but-curious) with respect to Sybil resistance and soundness, and fully malicious otherwise.
This is a common assumption, for the reasons stated above.
This leaves only a single goal of our malicious identity provider:
  breaking User-SP relationship anonymity.

\subsection{Oblivious Pseudorandom Functions}
\label{1:background:oprf}
In privacy-preserving computation, two parties may wish to compute a function they both input some data, but do not learn the other party's input.
This idealized functionality is realized by \emph{Oblivious Pseudo-Random Functions} (OPRFs).
An OPRF is a two-party protocol between a \enquote{client} and a \enquote{server} \cite{SoK:OPRF}.
The server contributes some secret key $k$, and the client contributes a value $X$.
After performing the protocol, the client learns $\OPRF_k(X)$, but learns nothing else about $\OPRF_k$ or $k$.
Meanwhile, the server learns no additional information; in particular, it does not learn $X$ or $\OPRF_k(X)$.

This very simple primitive is useful in a variety of applications;
 as a result, there are ongoing standardization efforts by the IRTF's Crypto Forum Research Group~\cite{DBLP:journals/rfc/rfc9497}.

In this work, we borrow operations from the specific class of Hashed Diffie-Hellman OPRFs over prime-order groups.
This is a widely-researched class of OPRFs \cite{SoK:OPRF}, which can be conceptualized as a sequence of three operations:

\begin{description}
\item[$\BLIND(X, r) := r \cdot X$:] A client function which uses a randomly-sampled $r \in_R \SCALARS$ to blind the input value $X$ for transmission to the server.
  If $X$ is not already a group element of the prime-order group, it is transformed to an element of $\POINTS$ by applying a suitable hash function.
\item[$\EVAL(A, k) := k \cdot A$:] A server function that evaluates the blinded input value to a blinded output value using the server's secret $k$.
If $k$ is not already of the appropriate form, it is first transformed into $\SCALARS$ by applying a hash function.
\item[$\UNBLIND(B, r) := r^{(-1)} \cdot B$:] A client function that removes the blinding from the output value to produce the final output.
  The result is then commonly hashed again. This, conveniently, also destroys any algebraic structure in the output.
\end{description}
The output $\OPRF_k(X) = \UNBLIND(\EVAL(\BLIND(X, r), k), r)$ \\is then $r^{(-1)} \cdot k \cdot r \cdot X = k \cdot X$.

We believe that it would be possible to generalize our approach to other types of OPRFs.
However, our security proof of pseudonym unlinkability in \Cref{1:analysis} relies on properties of this particular OPRF construction.
Thus, we leave this as potential future work and do not pursue it further here.

\subsection{BISON Pseudonym Derivation}
\label{1:protocol}

\begin{figure*}[t]
  \includegraphics{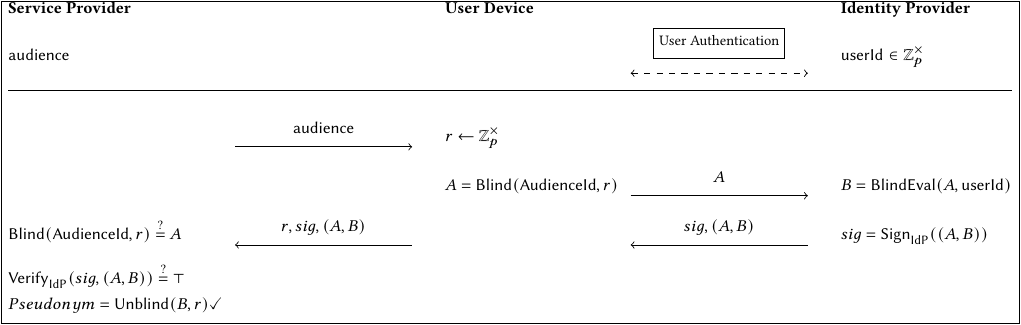}
  \caption{Simplified visual representation of the BISON protocol.}
  \label{fig:bison}
\end{figure*}

BISON applies OPRFs, introduced in the preceding section, to the three-party problem of pseudonym derivation.
The $\AUDSTR$, selected by the service provider and authenticated by the user device, is hashed to obtain $\AUDHASH \in \POINTS$, which serves as the client input $X$.
The $\UID \in \SCALARS$, only known to the identity provider, serves as the server input $k$.
We also add signatures to convince the service provider that the user truthfully performed the OPRF.
This allows the SP to trust the OPRF's output, which then serves as the user's BISON pseudonym.

The resulting BISON protocol is shown in full as \Cref{fig:prot1}.
We use the $\BLIND$, $\EVAL$, and $\UNBLIND$ operations of Hashed Diffie-Hellman OPRFs, as described in \Cref{1:background:oprf}.
A simplified visual representation is provided as \Cref{fig:bison}.

\begin{flushleft}
\captionof{protocol}{The BISON protocol as a sequence of steps.}
\label{fig:prot1}

  \vspace{3pt}\noindent\text{\textsc{Service Provider}}
  \begin{enumerate}[nosep]
    \item send desired $\AUDSTR$ to user
  \end{enumerate}

  \vspace{3pt}\noindent\text{\textsc{User Device}}
  \begin{enumerate}[nosep]
\setcounter{enumi}{1}
    \item receive $\AUDSTR$ from SP
    \item authenticate SP for $\AUDSTR$ \Comment{see \Cref{1:analysis:sp-authentication}}
    \item derive $\AUDHASH = H(\AUDSTR)$  %
    \item sample $\R \leftarrow \SCALARS$
    \item calculate blinded $\AUDHASH$: $\BLINDINPUT = \BLIND(\AUDHASH, \R)$
    \item send $\BLINDINPUT$ to IdP
  \end{enumerate}

  \vspace{3pt}\noindent\text{\textsc{Identity Provider}}
  \begin{enumerate}[nosep]
\setcounter{enumi}{7}
    \item receive $\BLINDINPUT$ from user
    \item authenticate user and obtain $\UID \in \SCALARS$
    \item calculate blinded pseudonym: $\BLINDOUTPUT = \EVAL(\BLINDINPUT, \UID)$
    \item sign $(\BLINDINPUT, \BLINDOUTPUT)$ to obtain signature $sig$
    \item send $(sig, (\BLINDINPUT, \BLINDOUTPUT))$ to user
  \end{enumerate}

  \vspace{3pt}\noindent\text{\textsc{User Device}}
  \begin{enumerate}[nosep]
\setcounter{enumi}{12}
    \item receive $(sig, (\BLINDINPUT, \BLINDOUTPUT))$ from IdP
    \item send $(\R, sig, (\BLINDINPUT, \BLINDOUTPUT))$ to SP
  \end{enumerate}

  \vspace{3pt}\noindent\text{\textsc{Service Provider}}
  \begin{enumerate}[nosep]
\setcounter{enumi}{14}
    \item receive $(\R, sig, (\BLINDINPUT, \BLINDOUTPUT))$ from user
    \item check $\BLINDINPUT = \BLIND(\AUDHASH, \R)$ \Comment{this validates $\R$}
    \item verify signature $sig$ over $(\BLINDINPUT, \BLINDOUTPUT)$ \Comment{see \Cref{1:analysis:idp-authentication}}
    \item derive pseudonym as $\UNBLIND(\BLINDOUTPUT, \R)$
  \end{enumerate}

\par\offinterlineskip\vskip3pt\hrulefill

\end{flushleft}

\section{BISON Evaluation}
\label{sec:evaluation}

\subsection{Protocol Analysis}
\label{1:analysis}

We now provide a security proof, and thus demonstrate that the BISON protocol fulfills the requirements of \Cref{1:goals}.

We note that we intentionally do not include protections against message replay in the core protocol described above.
This allows for easier integration into existing authentication protocols.
Such protocols must already mitigate message replay; work which we would duplicate.
We argue this point further in \Cref{1:analysis:replay}.
If desired, the BISON protocol can be augmented to include such protection.

We also do not mandate a particular method of authenticating service providers for a given $\AUDSTR$.
This is because different methods will be appropriate depending on the context in which BISON is used.
We also discuss this requirement, and some possible solutions, further in \Cref{1:analysis:sp-authentication}.

Finally, we do not specify how the identity provider authenticates the user.
This is common for authentication protocols.

\myparagraph{Pseudonym Unlinkability}
Two service providers belonging to different audiences each perform an authentication process, and obtain pseudonyms $\UID_1 \cdot \AUDHASH_A$ and $\UID_2 \cdot \AUDHASH_B$ respectively.
Since the user has performed context-specific steps\footnote{see \Cref{1:analysis:sp-authentication}} to authenticate each SP for their $\AUDHASH$, we know that $\AUDHASH_A \neq \AUDHASH_B$.
We show that the two service providers cannot decide whether $\UID_1 = \UID_2$.

To do this, we first observe that the $\AUDHASH$ values are obtained by applying a cryptographic hash function to the $\AUDSTR$s.
Therefore, we can model $\AUDHASH_A$ and $\AUDHASH_B$ as being sampled at random from $\POINTS$.
Additionally, we note that $\UID_1$ and $\UID_2$ were also sampled randomly from $\SCALARS$ with sufficient entropy, as discussed in \Cref{1:discussion:entropy}.
This allows us to state \Cref{thm:unlinkability}.

\begin{lemma}
\label{thm:unlinkability}
If, given the quadruple $(\AUDHASH_A, \allowbreak\AUDHASH_B, \allowbreak\UID_1 \cdot \AUDHASH_A, \allowbreak\UID_2 \cdot \AUDHASH_B)$, an attacker is capable of deciding whether $\UID_1 = \UID_2$, that attacker can also decide the Decisional Diffie-Hellman Problem (DDHP) \cite{DDHP}.
\end{lemma}

\begin{proof}
We are given an instance of DDHP, which is a tuple $(a \cdot G,\allowbreak b \cdot G,\allowbreak c \cdot G)$ for $(a,b,c) \in_R \SCALARS$ and generator $G \in \POINTS$.
To decide DDHP, we need to decide whether $c \cdot G = (a \cdot b) \cdot G$.
We map this to an instance of our problem as follows: let $\AUDHASH_A := G$, $\AUDHASH_B := a \cdot G$, $\UID_1 \cdot \AUDHASH_A := b \cdot G$, $\UID_2 \cdot \AUDHASH_B := c \cdot G$.
It immediately follows that $\UID_1 = b$.
As per our assumption, our attacker can now decide whether $\UID_2 = \UID_1$.
If $\UID_2 = \UID_1$, then $c \cdot G = \UID_2 \cdot \AUDHASH_B = \UID_1 \cdot \AUDHASH_B = b \cdot \AUDHASH_B = b \cdot a \cdot G = (a \cdot b) \cdot G$.
Furthermore, due to $G$'s nature as a generator, $\UID_2 \neq \UID_1$ similarly implies $c \cdot G \neq (a \cdot b) \cdot G$.

Thus, our attacker can decide DDHP.
\end{proof}

Note that we assume that deciding the DDHP on $\POINTSGROUP$ to be hard.
This extends to cases of $k > 2$ different service providers, reducing to the Generalized Decisional Diffie-Hellman problem~\cite{DBLP:conf/icics/BaoDZ03}.

\myparagraph{User-SP Relationship Anonymity}
The authentication process results in the identity provider learning $\R \cdot \AUDHASH$ and $\UID \cdot \R \cdot \AUDHASH$.
The identity provider knows $\UID$, and we also assume it has a full list of potential $\AUDHASH$s to test, small enough to be reasonably iterated.

Nevertheless, this is not a concern.

\begin{lemma}
\label{thm:anonymity}
If $\R \in_R \SCALARS$, then $\forall X \in \POINTS: \mathbb{P}(\R \cdot \AUDHASH = X) = \frac{1}{\ORDER-1}$ regardless of choice of $\AUDHASH \in \POINTS$.
\end{lemma}

\begin{proof}
Recall that we assumed $\POINTSGROUP$ to have prime order.
In a prime-order group, any non-identity element is a generator of the group.
Since $\AUDHASH$ is a non-identity element, $\AUDHASH$ generates $\POINTSGROUP$.
Therefore, $\star_\AUDHASH: \SCALARS \rightarrow \POINTS$ defined by $x \mapsto x \cdot \AUDHASH$ is a bijection.
$\R \in_R \SCALARS$ was chosen uniformly at random, i.e., $\forall x \in \SCALARS: \mathbb{P}(\R = x) = \frac{1}{\ORDER-1}$.
As a result, $\forall X \in \POINTS: \mathbb{P}(\star_\AUDHASH(\R) = X) = \frac{1}{\ORDER-1}$.
This is true independently of the choice of $\AUDHASH$.
Now recall that $\star_\AUDHASH(\R) = \R \cdot \AUDHASH$.
\end{proof}

To summarize: due to the unknown, random $\R \in_R \SCALARS$, the blinded input $\R \cdot \AUDHASH$ contains no information about $\AUDHASH$.
It is thus impossible for the identity provider to gain any information from the blinded values, as long as they do not learn $\R$.

\myparagraph{Sybil Resistance}
\label{1:analysis:sybil}
The authentication process results in the service provider learning $(\R \cdot \AUDHASH, \UID \cdot \R \cdot \AUDHASH)$, signed by the identity provider and therefore trustworthy.
They then receive an untrusted claimed value of $\R$ from the user, which we will refer to as $\R'$.
Using these values , the user's pseudonym is derived as $(\R')^{-1} \cdot \UID \cdot \R \cdot \AUDHASH$.

Before doing so, they verify that $\R \cdot \AUDHASH$ (as signed by the IdP) equals $\R' \cdot \AUDHASH$ (for $\R'$ as claimed by the user).
We show that this is sufficient.

\begin{lemma}
\label{thm:soundness}
For all $\R, \R' \in \SCALARS$, if $\R \cdot \AUDHASH = \R' \cdot \AUDHASH$ holds, it follows that $\R = \R'$.
\end{lemma}

\begin{proof}
Following the argument used for \Cref{thm:anonymity} above, recall that $\star_\AUDHASH: x \mapsto x \cdot \AUDHASH$ is a bijection from $\SCALARS$ to $\POINTS$.
It follows that $\star_\AUDHASH(\R) = \R \cdot \AUDHASH = \R' \cdot \AUDHASH = \star_\AUDHASH(\R')$ if and only if $\R = \R'$.
\end{proof}
As a result, $(\R')^{-1} = (\R)^{-1}$, and thus we can be sure that the learned pseudonym is actually $\UID \cdot \AUDHASH$.
This value is constant across authentication processes for this fixed $\AUDHASH$, as we assume that the identity provider honestly chooses the correct $\UID$ per \Cref{1:threat-model}.

\myparagraph{No persistent user device state}
Our protocol does not require any state on the user's device.
We consider this to be self-evident from the description in \Cref{1:protocol}.

\myparagraph{Soundness \& Validity}
During the authentication process, the service provider obtains the tuple $( \BLINDINPUT, \BLINDOUTPUT )$, signed by the identity provider.
We assume%
 \footnote{see \Cref{1:analysis:external}}
 that the service provider can validate that this was indeed signed by the genuine identity provider;
  a malicious user cannot forge such a signed tuple.

Furthermore, we also assume\footnotemark[\value{footnote}] that the service provider can validate that the response presented is fresh, and intended to be part of the ongoing authentication process.
Therefore, even a malicious user can only obtain a valid tuple in real time from the genuine identity provider.
Per \Cref{1:threat-model}, we assume that the genuine identity provider only issues such tuples after successful user authentication for $\UID$.
As part of this process, it can also validate that the user's account is not suspended or deleted.

As shown in \Cref{thm:soundness}, the service provider's validation of $\BLINDINPUT = \R \cdot \AUDHASH$ ensures that the result of the derivation is indeed the user's BISON pseudonym, $\UID \cdot \AUDHASH$.
Therefore, a malicious user cannot produce an authentication response for any $\UID$ except their own.

\subsection{External Considerations}
\label{1:analysis:external}

For BISON to meet our stated goals, we rely on the \enquote{host} protocol, which embeds BISON derivation, to provide certain guarantees.
These are guarantees that are either context-specific, or already a necessary part of existing authentication protocols.
Thus, we choose not to mandate a particular way of achieving them.
Instead, we merely list these requirements here, and expect the host protocol to ensure they are met.

\myparagraph{Replay}
\label{1:analysis:replay}
It may be possible for a malicious user to have knowledge of signed $( \BLINDINPUT, \BLINDOUTPUT )$ tuples from a previous authentication process;
  for example, they may have previously compromised a legitimate user's account, their previously-valid account may have been suspended by the identity provider, or they may have compromised a user's device.
The host protocol needs to ensure that a malicious user cannot \emph{replay} such a previously-issued tuple to the service provider, instead of obtaining a fresh one from the identity provider.
This attack would allow them to complete the authentication process without needing to authenticate to the identity provider.

Replay attacks are not unique to BISON, but are a challenge that any authentication protocol needs to solve.
Therefore, any host protocol embedding BISON derivation will already have some means of guaranteeing response freshness.
This is commonly done by including a one-time redemption step~\cite{OIDC}, or adding a challenge nonce to the message that must be signed by the IdP~\cite{webauthn}.
Adding such a step to the core BISON protocol would be redundant.

We thus decided not to add additional replay safeguards to the core BISON protocol, which we described in \Cref{1:protocol} above.
However, if replay protection as part of the BISON derivation is desired, it can easily be added as follows:
  have the service provider sample blinding randomness $\R$, and provide it to the user device.
The user device then does not sample its own randomness, but is instead required to use the randomness provided by the service provider.
This randomness, and therefore the blinded audience identifier $\BLINDINPUT$, will be different and unpredictable for each authentication process.
Since $\BLINDINPUT$ is part of the identity provider's signed response, this forces the signed response to be freshly generated as part of this particular authentication process, and prevents message replay.

\myparagraph{IdP Authentication}
\label{1:analysis:idp-authentication}
The BISON protocol has the identity provider sign an $( \BLINDINPUT, \BLINDOUTPUT )$ tuple, which is forwarded by the user to the service provider.
The service provider then needs to somehow be able to validate the signature over this tuple.

How the necessary key material is obtained, and how its authenticity is assured, is context-specific.
This is a challenge that is not unique to BISON.
Instead, it is the same challenge that any other delegated authentication protocol needs to solve.
Thus, once again, any host protocol embedding BISON derivation will already have the necessary mechanisms in place.

Additionally, care must be taken to tie the signed data to the underlying protocol's authentication process.
An attacker must not be able to take a BISON tuple from one authentication process, and transplant it onto a second parallel authentication process.
The means by which this binding is achieved are protocol-specific;
  to give an example, in OpenID Connect, $\BLINDINPUT$ and $\BLINDOUTPUT$ could be embedded into the monolithic \enquote{ID token} signed by the identity provider.
This concrete instantiation will be discussed further in \Cref{2:oidc-extension}.

\myparagraph{SP Authentication}
\label{1:analysis:sp-authentication}
The BISON protocol makes the user device responsible for verifying that the service provider is authorized for the requested $\AUDSTR$.
Failure to do so would allow a monster-in-the-middle (MitM) attack;
  a malicious entity could forward the genuine service provider's request to the user, and the user's response to the service provider.
The service provider would believe that the malicious entity has successfully authenticated as that user.

Additionally, we also note that if multiple service providers are permitted to share an $\AUDSTR$, they are able to link users' requests (by design).
There is potential for them to perform a MitM attack on connections to each other.
In contexts where such SPs do not always implicitly trust each other, the authentication process should also be bound to a particular service provider instance.%
  \footnote{This can be done by, e.g., including a randomized commitment to the SP in the data signed by the IdP. This is the approach we use in our demonstrator in \Cref{2:implementation}.}

How a service provider is authenticated is context-specific.
Therefore, the BISON protocol does not mandate any particular method of $\AUDSTR$ authentication.
This may be derived from an existing trust scheme, such as a DNS hostname listed in a Web PKI X.509 certificate;\footnote{This is the approach we use in \Cref{2:implementation}.}
  or a new context-specific trust scheme, such as certificates issued by the identity provider or some central authority.

Different forms of authentication carry different guarantees.
For example, deriving it from the hostname does not allow the identity provider control over which service providers use its services;
  while requiring certificates issued by the identity provider, with appropriate revocation mechanisms such as OCSP stapling in place, would, at the cost of additional complexity.
The appropriate method therefore depends on the context in which BISON is used.

We also note that BISON does not allow the identity provider to independently verify the service provider's authorization.
It must rely on the user device to faithfully perform this validation.

Finally, we caution that authentication must be tied to information authenticated by an encapsulating transport protocol, such as TLS.
If appropriate care is not taken here, the MitM attack originally outlined may still be possible.

\subsection{Protocol Discussion}
\label{1:discussion}

In this section, we highlight the risks of predictably-chosen $\UID$ values.
We also discuss how a user could verify that the IdP is not misbehaving.

\myparagraph{Collusion between IdP and SP}
\label{1:discussion:unmasking}
The BISON pseudonym obtained by service providers is $\UID \cdot \AUDHASH$.
The identity provider never learns this pseudonym.

But -- what happens if a SP provides IdP with a pseudonym and the corresponding $\AUDHASH$?
The IdP has a list of their registered users and the corresponding $\UID$s;
  they can iterate over them, perform trial calculations, and easily determine the $\UID$ underlying the pseudonym.

We show that this is not a limitation of BISON in particular.
Instead, \emph{any} protocol $\mathcal{P}$ with stable pseudonyms, without persistent user device state, which allows the IdP to authenticate users through arbitrary unspecified means, must exhibit this limitation.

Assume that a service provider $A$ colludes with the identity provider, and provides them with pseudonym $P_{X,A}$ for unknown user $X$.
The identity provider by necessity has a list of possible values of $X$.
For any candidate user $X'$, the identity provider can now run $\mathcal{P}$ locally, impersonating both the service provider $A$ and user $X'$.
This must be possible -- $X'$ has no persistent state, and the identity provider can always \enquote{authenticate} itself for $X'$, since $\mathcal{P}$ places no constraints on the authentication method(s).
As a result, the identity provider therefore learns $P_{X',A}$.
Repeating this process for each candidate user $X'$ will inevitably find $X$, the only user where $P_{X',A} = P_{X,A}$.

Therefore, susceptibility to IdP-SP collusion is a necessary limitation of any protocol meeting BISON's goals.

\myparagraph{Entropy of global user IDs}
\label{1:discussion:entropy}
As shown in \Cref{1:analysis}, correlating two BISON pseudonyms would usually require deciding the Decisional Diffie-Hellman Problem (DDHP).
Similarly, calculating the $\UID$ from the BISON pseudonym ($\UID \cdot \AUDHASH$) requires solving the Discrete Logarithm Problem (DLP).
We assume both of these problems to be hard.

However, there is a fundamental assumption baked into these statements -- that the $\UID$ is sufficiently unpredictable to be essentially random.
This is not the case if, for example, a simple auto-incrementing database key is used as the $\UID$;
  naïve iteration over the integers could then allow the SP, knowing $\AUDSTR$, to determine the $\UID$.

Therefore, the IdP must ensure that $\UID$ carries sufficient entropy.
One straightforward approach is sampling a random byte string of length similar to that of the group order;
  though depending on the context, others may be preferred.

This issue is not unique to BISON; it is shared by any similar pseudonym derivation scheme.

\myparagraph{Verifiability}
\label{1:discussion:verifiability}
In BISON, the user relies on the IdP using the correct $\UID$ in the pseudonym derivation.
The user can, to some extent, verify that the IdP is behaving correctly in this regard:
  they can remember their pseudonym for any $\AUDSTR$ they have previously authenticated to, and (see \Cref{2:discussion}) can verify that the IdP-provided tuple produces that pseudonym.
Note that the IdP learns no information regarding the $\AUDSTR$ involved with any given authentication process;
  in particular, the IdP cannot know whether this is an $\AUDSTR$ the user has previously authenticated to, rendering malfeasance detectable.

This ability could be extended to previously-unused $\AUDSTR$s by borrowing from so-called \emph{verifiable} OPRFs~\cite{DBLP:journals/rfc/rfc9497}.
Here, the server (i.e., the IdP) calculates a noninteractive zero-knowledge proof of discrete logarithm equivalence;
  in BISON terms, they could produce a proof that the multiplications $\UID \cdot \R \cdot \AUDHASH$ and $\UID \cdot G$ used the same scalar multiplier $\UID$ without disclosing it.
The user device could verify this proof, and could thus ensure that the IdP always uses the same $\UID$ for the same user.
However, this also only provides verifiability \emph{by the user}; the same scenario already covered by the argument we made in the previous paragraph.

In particular, it still does not prevent a malicious user and identity provider colluding against a service provider.
Therefore, we did not adopt this idea for BISON as described in this work.

\subsection{Performance Evaluation}
\label{1:evaluation}

A full BISON derivation process consists of sampling of some randomness $\R$ and the procedures $\BLIND$, $\EVAL$, $\BLIND$ (again, to verify), and $\UNBLIND$.
When using a prime-order OPRF over an elliptic curve group, this process requires 4 elliptic-curve scalar-point multiplications and 4 hash function evaluations.

When applied in an authentication protocol (cf. \Cref{2:section}), SP, User, and IdP need to perform BISON derivation in addition to the established authentication process and network communication.
To maintain usability, it is thus important that BISON does not slow down the process.
To evaluate the performance impact of BISON, we develop a proof of concept implementation in Kotlin.
We instantiate BISON on the ristretto255 curve using the SHA-512 hash function and benchmark the implementation using the JMH benchmarking framework.\footnote{\url{https://github.com/openjdk/jmh}}
We run the benchmarks on a standard office laptop -- a Lenovo ThinkPad T14 G2, with a 2.8 GHz Intel Core i7-1165G7 processor, and typical background load.

On our reference machine, the full BISON derivation process takes $\approx 3$ milliseconds.
We took no particular steps to improve the benchmark's performance.
Therefore, based on these results, we conclude that usage of BISON derivation has a completely negligible performance impact.

\section{BISON-augmented OpenID Connect}
\label{2:section}

In this second part of our work, we demonstrate the practical feasibility of BISON derivation.
To do this, we specify an extension to the widely-used OpenID Connect authentication protocol.
This enables it to support BISON pseudonyms.
If the user device is unaware of or does not support BISON, the extension transparently falls back to \enquote{standard} OpenID Connect.

\subsection{Background}
\label{2:background}

\subsubsection{OpenID Connect (OIDC)}

OpenID Connect is one of the most widely used authentication protocols on the current internet.
Building on the underlying OAuth 2.0 authorization protocol, it adds an authentication layer, allowing service providers to obtain a user identifier suitable for re-authentication \cite{OIDC,DBLP:conf/eurosp/MainkaMSW17}.
It has vast adoption: Amazon, Apple, DropBox, Facebook, Google, IBM, Linkedin, Microsoft, and Yahoo all implement OpenID Connect in their respective identity providers \cite{SHARIF2022103097, AppleOIDCDocs}.

The goal of an OIDC authentication process is for the service provider to obtain a so-called \emph{ID Token}.
This token is a signed, JSON-encoded set of key-value pairs, called \emph{claims}, about the user.
One such claim is the user's \emph{subject identifier} (\texttt{sub} claim).
The ID Token is signed by the identity provider, and its integrity can be verified using the identity provider's previously-published public key.

Conceptually, the process starts with the service provider, often reacting to a user action, creating an \emph{Authentication Request}.
As part of this request, the service provider identifies itself (using a \texttt{client\_id}), specifies where the user should be returned to (the \texttt{redirect\_uri}), and commonly also includes a \texttt{nonce} to prevent message replay.
It then typically issues an HTTP redirect, or submits an HTML form, making the user device contact the selected identity provider with the request.

In reaction to the request, the identity provider prompts the user for any necessary authentication factors to verify their identity.
It then prompts them for consent to forward their information to the requesting service provider.
If consent is granted, the identity provider issues the aforementioned ID Token.
The token includes the user's subject identifier, alongside information used to mitigate various attacks -- such as the IdP (\texttt{iss}) and SP (\texttt{aud}) identifiers, expiry time (\texttt{exp}), and time of issuance (\texttt{iat}).
If requested, it also includes the challenge nonce (\texttt{nonce}).
Once the ID Token has been created, the user device is redirected back to the service provider.

OpenID Connect supports different \emph{operation flows}, which change how exactly the service provider now obtains the signed ID Token.
In \emph{implicit} flow, the ID Token is immediately included when redirecting back to the service provider.
The service provider's identity is only validated implicitly by the user device, as the user is redirected back to a known HTTPS-enabled URI.

By contrast, in \emph{authorization code} flow, a one-time-use code is included when redirecting back to the service provider.
The service provider can then redeem this one-time-use code for the ID Token via a back channel to the identity provider.
This redemption process often also includes explicit verification of the service provider's identity by the identity provider, via a pre-shared \enquote{client secret}.

While our proof of concept uses the implicit flow, it is possible to use BISON in authorization code flow.
This requires solving additional challenges;
  and BISON already provides service provider authentication, which is the main benefit of authorization code flow.
Therefore, we consider the implicit flow to be the natural choice when using BISON.
We discuss this further in \Cref{2:discussion}.

\subsubsection{Pairwise Pseudonymous Identifiers (PPIDs)}

The OpenID Connect specification~\cite{OIDC} allows for the use of \emph{pairwise pseudonymous identifiers} (PPIDs).
Here, the identity provider derives a pseudonymous identifier for a particular audience instead of revealing a global identifier.
For example, Apple's \enquote{Sign in with Apple} OpenID provider generates a different pseudonymous identifier based on the service provider's associated Apple Developer Account~\cite{AppleOIDCDocs}.

The particular derivation method, and audience delineation, are left up to the identity provider.
Commonly, a hash digest of the user's global identifier concatenated with some audience identifier is calculated and used as the user's PPID~\cite[Sec 8.1]{OIDC}.
Traditional PPID derivation envisions the IdP knowing which audience that the user is authenticating to.
This enshrines the need for the IdP to learn sensitive association data with each login;
 it cannot coexist with User-SP relationship anonymity towards the identity provider.

\subsubsection{FedCM \& Browser-augmented OIDC}
\label{2:background:fedcm}

Traditional OpenID Connect authentication flows, as described above, assume, and work around, a protocol-unaware user device.
This is achieved by leveraging general-purpose technologies, such as HTTP redirects and HTML form submissions, to make the user's browser forward opaque information between logically unrelated remote servers.

Recently, many such technologies have come under scrutiny.
This comes in the wake of web browsers limiting the traditional ways of tracking user behavior across disparate web origins, such as third-party cookies, which have long been misused by the online advertising sector.
In reaction, privacy adversaries have explored new ways of associating user interactions;
  this includes the use of message-carrying redirects to make the browser identify itself to an advertising tracker.
In response, Browser manufacturers propose limiting such \emph{stateful redirects}.
This would eliminate the technology that OpenID Connect, and similar use cases, depend on.

The current W3C proposal to solve this conundrum is the \emph{FedCM API}~\cite{fedcm,jannettsok}.
It envisions browsers being conceptually aware of authentication processes, and taking an active role in negotiating user authentication.
This represents a paradigm shift away from the traditional OIDC-unaware user device.

\subsection{BISON OIDC Extension}
\label{2:oidc-extension}

We build on the existing OIDC infrastructure by specifying BISON pseudonym derivation as an opt-in PPID derivation method.

Our extension consists of two sets of changes.
The first set supports compatibility autodiscovery and transparent fallback.
\rom{1}. We add a new optional \texttt{pairwise\_subject\_types} array to the identity provider's existing OpenID Connect Discovery metadata~\cite{oidc-discovery}.
This array advertises the cryptographic derivation methods supported. We define \texttt{bison} as one possible value for this array.
\rom{2}. The service provider may opt into any of the derivation methods listed by specifying the chosen method(s) in a \texttt{pairwise\_subject\_types} array in the authentication request.
This request is otherwise a well-formed standard OIDC request. Therefore, in a protocol-unaware browser, it will be processed normally by the identity provider.
\rom{3}. The identity provider includes the used derivation method (if any) as a \texttt{pairwise\_subject\_type} claim in the signed ID Token.

Our second set of changes integrates BISON derivation into OpenID Connect.
\rom{4}. If the service provider opts into BISON derivation, a supporting user device modifies the OIDC authentication request in three ways, and also sets \texttt{pairwise\_subject\_type=bison}.
  (a). It replaces OIDC's existing service provider identifier field (\texttt{client\_id}) with BISON's $\BLIND$-ed audience identifier;
  (b). it replaces the return location (\texttt{redirect\_uri}) with a constant value\footnote{Our demonstrator uses the invalid domain \texttt{https://anonymous.invalid/bison}.}, preventing disclosure;
  (c). and it replaces the SP's chosen \texttt{nonce} with the hash digest of the current secure origin\footnote{This is necessary because the current origin may be different from the $\AUDSTR$. See \Cref{2:analysis} for details on $\AUDSTR$ authentication.}
     and the \texttt{nonce}.
\rom{5}. If BISON derivation has been requested, the identity provider $\EVAL$-s the provided \texttt{client\_id} value alongside the user's global identifier.
The resulting blinded pseudonym is stored in the existing subject field (\texttt{sub} claim) of the resulting signed JWT, while the input blinded audience identifier is stored in the existing audience field (\texttt{aud} claim).
\rom{6}. When redirecting back to the service provider, the user device adds the randomly selected blind as an (unsigned) \texttt{blind} parameter.
\rom{7}. After the service provider has verified the signed ID Token, it performs BISON derivation.
  (a). It re-calculates $\BLIND$ and validates that \texttt{aud} is correct.
  (b). It verifies that the value of \texttt{nonce} is as expected.
  (c). It derives the user's BISON pseudonym by $\UNBLIND$-ing \texttt{sub}.

\subsection{BISON-OIDC Analysis}
\label{2:analysis}
We show that our integration meets the host protocol security criteria described in \Cref{1:analysis:external}.
As a result, we conclude that BISON-OIDC provides the privacy guarantees described in \Cref{1:goals}.

\myparagraph{Replay}
In standard OpenID Connect, replay protection takes the form of a challenge nonce;
  a random string value that is chosen by the service provider and sent along with the authentication request.
When the identity provider creates the signed ID token for this authentication process, it includes the provided nonce.
This guarantees that the signed ID token was created in response to this particular request, as the nonce was not known beforehand.

In BISON-OIDC, we maintain this basic idea.
Additionally, we mitigate the same-audience MitM attack outlined in \Cref{1:analysis:external}.
We do this by taking the random nonce requested from the SP, and adding the current user-facing secure web origin -- not the $\AUDSTR$.
The result is then hashed and base64url-encoded.
This forms the \enquote{nonce} that is sent to the IdP for signing.
When the SP verifies the ID token, it can also verify the expected nonce value.

Any attempt at message replay would fail this verification, as the underlying SP nonce used as a hash input is different.
Meanwhile, any attempt at a same-audience MitM attack would also fail this verification, as the user-facing origin would not be as expected.

\myparagraph{IdP Authentication}
In OpenID connect, the ID token is signed by the identity provider. (cf. \Cref{2:background})
The public key can be obtained by server providers through a variety of means.
BISON-OIDC does not require any special consideration in this regard.

\myparagraph{SP Authentication}
We bind audience identifier authorization to the underlying HTTPS browsing context.
This is the same approach used by existing web standards, such as Web Authentication~\cite{webauthn}.
We limit BISON-OIDC secure browsing contexts, such as top-level windows serving content via HTTPS.\footnote{For a more detailed definition of this term of art, we refer to \url{https://w3c.github.io/webappsec-secure-contexts}.}
This implies that the browser has validated the current page's TLS certificate using its internal trust store.
Thus, under the standard assumption that the Web PKI can be trusted, the current page origin is trusted information.
We use this fact for $\AUDSTR$ validation.

We define acceptable $\AUDSTR$ values to be either the current page origin, or a registrable domain suffix of the current page origin.%
\footnote{For a detailed definition, we refer to \url{https://html.spec.whatwg.org/multipage/browsers.html}.}
This mirrors the limitation placed on the relying party identifier field in Web Authentication~\cite{webauthn}.
For example, \path{login.example.com} and \path{app.example.com} may derive a shared pseudonym by using the \path{example.com} $\AUDSTR$.
However, \path{login.example.com} and \path{login.unrelated.com} cannot derive a shared pseudonym, as \path{.com} is not registrable.
Additionally, we limit the return address to being on the same web origin as the current page origin.
This achieves audience identifier authentication in the standard web trust model.

\subsection{Implementation}
\label{2:implementation}

We present a full implementation of the OIDC extension described in \Cref{2:oidc-extension}.
It takes the form of a BISON-capable service provider/identity provider pair, as well as an extension making the Firefox browser protocol-aware.
The source code is available on GitHub.\footnote{\THEREPO \label{fn:therepo}}
We also host a public demonstrator instance.\footnote{\url{https://bison.grazing.website} \label{fn:thedemo}}  %
We describe key details here, and leave a full description to \Cref{sec:appendix:impldetails}.

We instantiate BISON-OIDC as follows.
For our prime-order group, we choose the \emph{ristretto255} elliptic curve group~\cite{ristretto255}.
As our hash function, we choose SHA512.
We note that our choices match the parameters of an OPRF instantiation in RFC 9497~\cite{DBLP:journals/rfc/rfc9497}, and discuss BISON instantiations further in \Cref{2:discussion}.

Our implementation models a minimal authentication process using OIDC form-post implicit flow.
The identity provider performs mock \enquote{authentication} by allowing a choice between one of three pre-defined identities.
Users visiting the service provider are redirected to the identity provider to log in.
This makes a BISON-OIDC authentication request.
If the user has not installed the Firefox browser extension, this transparently falls back to \enquote{regular} OIDC.

\begin{figure}
\includegraphics[width=\linewidth]{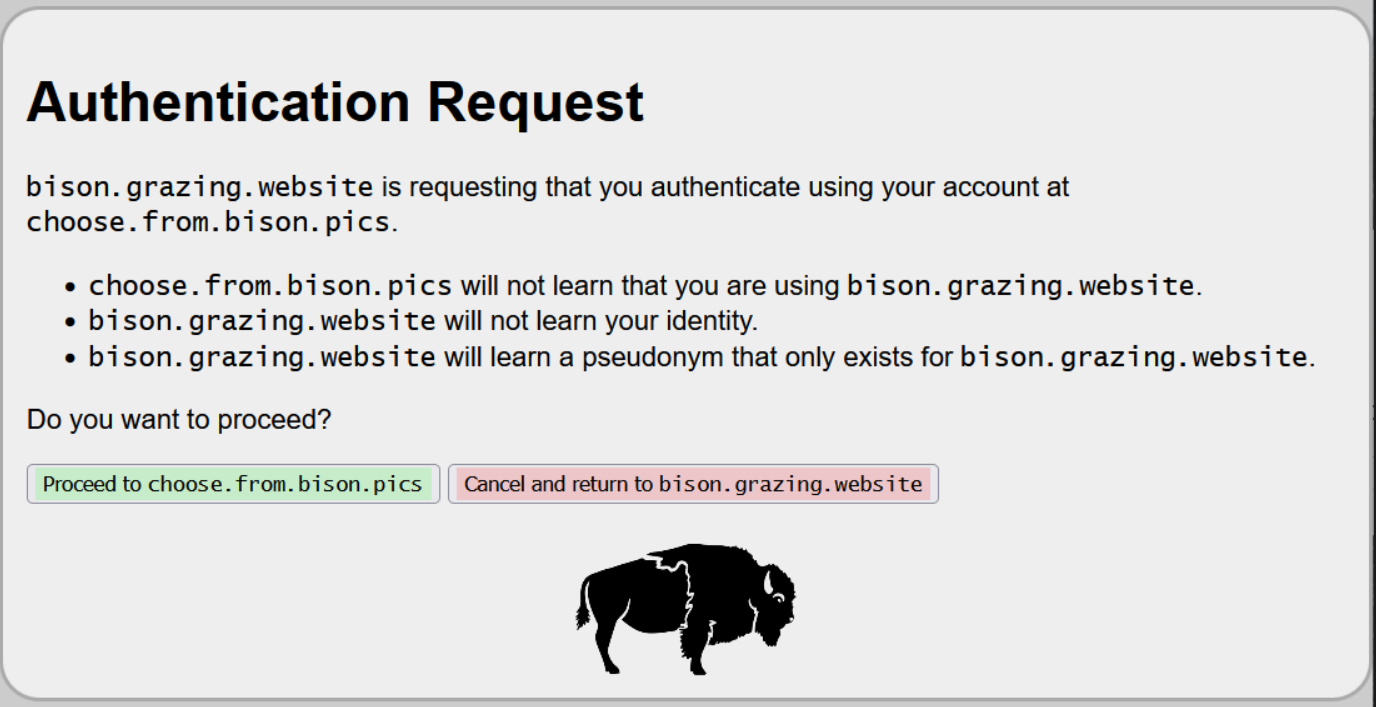}
\caption{The BISON consent dialog.}
\label{fig:authn-dialog}
\end{figure}

If the user has installed the extension, the authentication request is recognized as such.
Our logic then intercepts it and prompts the user for consent.
When doing so, it displays the service provider's origin, as well as the requested $\AUDSTR$.
This is depicted in \Cref{fig:authn-dialog}.
If consent is given, the authentication request is modified as described in \Cref{2:oidc-extension}.
The \texttt{redirect\_uri} is set to the constant value \texttt{https://anonymous.invalid/bison}.
As \texttt{invalid} is a reserved TLD, this domain is guaranteed not to exist~\cite{DBLP:journals/rfc/rfc2606}.

When the identity provider redirects the user back to this invalid domain, this is intercepted by the extension.
It attaches the \texttt{blind} and redirects back to the original \texttt{redirect\_uri}.

\subsection{Evaluation}
\label{2:evaluation}

We compare our OpenID Connect extension against a \enquote{traditional} OIDC-based authentication flow.
For overhead induced by the abstract BISON protocol, we refer to our prior evaluation in \Cref{1:evaluation}.
Instead, we focus on differences unique to browser-based login.

We consider the changed user experience of BISON-augmented authentication.
In BISON, the identity provider is no longer aware of the service provider's identity.
Therefore, its UX also cannot ask the user's consent for the authentication.
Instead, this step needs to be done by the browser.
In our proof of concept implementation, we include a dedicated dialog, shown in \Cref{fig:authn-dialog}, for this purpose.

This dialog is displayed before the user is redirected to the identity provider, and presents an additional step compared to traditional, non-BISON OIDC.
However, in a traditional UX flow, after the user device authenticates to the IdP, the IdP would then commonly present an additional page requesting user consent.
This additional page is unnecessary in BISON, as consent has already been obtained by the user device.
Therefore, we conclude that BISON-augmented OIDC does not require additional user interaction compared to traditional OIDC;
  it merely changes the ordering, shifting user consent to be before authentication.

Additionally, we note that BISON-augmented OIDC, in contrast to unaugmented OIDC, now requires a protocol-aware user device.
We discuss this further in \Cref{2:discussion}.

\subsection{Discussion}
\label{2:discussion}

This section lists some additional subtle distinctions between BISON-augmented and unaugmented OIDC.
We address these differences' impact on the practical use of BISON, and also motivate our choice of cryptographic parameters.

\myparagraph{Protocol Awareness in Browsers}
One crucial benefit of OpenID Connect, which has contributed to its widespread adoption, is that it does not require the user device to be aware of its existence.
BISON-augmented OIDC does not offer this benefit.
It requires the web browser to take an active role:
  verify the $\AUDSTR$, perform blinding operations, and obtain user consent using built-in dialog windows.
Thus, the web browser must be protocol-aware.

We argue that a protocol-aware browser is a necessary requirement for achieving relationship anonymity, pseudonym unlinkability, and Sybil resistance simultaneously.
To show this, we model authentication as a two-party protocol between a service provider and an identity provider;
  they use a protocol-unaware browser as a communication channel.
Assume that we have two distinct service providers, A and B, and a user, X.
The identity provider now receives an authentication request AuthReq.
At this point, AuthReq must not allow the identity provider to determine whether it originated from A or B;
  if this were possible, we would lose relationship anonymity.
After some user interaction to authenticate X, some data is returned to the originator, which results in pseudonym P.
Since we assume Sybil resistance, P needs to be stable across authentication requests.
However, recall that AuthReq might have originated from either A or B; therefore, it cannot provide pseudonym unlinkability.

A few years ago, the need for a protocol-aware browser\footnote{Or a browser extension, as in our demonstrator; we find this to be impractical for widespread adoption.} would have relegated our work chiefly into the realm of purely academic wish-casting;
  a castle on a cloud, never to be realized in practice.
But the internet moves along, and so does practicality.
Browsers are moving towards standardizing federated authentication;
 using semantics to decide between legitimate and illegitimate uses of technology.
We described these efforts in \Cref{2:background:fedcm}.

In this environment, we believe the requirement for active browser engagement in authentication is no longer disqualifying.
We believe the time has come for browsers to take an active role in safeguarding their users' privacy -- and BISON is the tool to do it with.

\myparagraph{Backwards Compatibility}
Any kind of extension to an existing protocol will initially be unsupported by already-deployed implementations.
For this reason, BISON-OIDC transparently falls back to \enquote{normal} OIDC.
If the IdP advertises support and the SP recognizes this, the SP can opt-in in its request (\texttt{pairwise\_subject\_types=bison}).
This request, when encountered by a protocol-aware browser, will be treated specially, with the browser performing blinding.
If the browser is unaware, though, this request is still a well-formed OIDC request, which the IdP will process normally without performing BISON operations.
Thus, existing functionality is not impeded by adopting BISON-OIDC.

\myparagraph{Choice of OIDC flow}
OpenID Connect envisions different ways (called \enquote{flows}, cf. \Cref{2:background}) to transmit the signed \emph{ID Token} from the identity provider to the service provider.

In our PoC, we use \emph{implicit} flow, even though \emph{authorization code} flow is common in practice.
This has multiple reasons.

Most obviously, authorization code flow uses a \enquote{back-channel}, a direct connection between the SP and IdP, to exchange the ID Token.
This back-channel presents an obvious challenge to our relationship anonymity goal;
 the ID Token corresponds to a particular user's login process, while the entity requesting it is the service provider they logged in at.
To maintain relationship anonymity, the service provider would thus need to connect to the identity provider without revealing its identity.
While such an anonymous connection is possible~\cite{DBLP:conf/uss/DingledineMS04}, realizing this in practice is not trivial.

We now consider the primary benefit of authorization code flow compared to implicit flow:
 service provider authentication.
When redeeming the authorization code for the ID Token, the service provider commonly provides a pre-shared secret to authenticate itself.
This prevents a malicious user, or service provider impersonator, from obtaining the ID Token and identifying information.
In BISON, however, the service provider cannot authenticate to the identity provider -- this would be antithetical to an anonymous connection to the identity provider.
Instead, \emph{the user} authenticates the service provider for the given $\AUDSTR$; authentication towards the identity provider is thus rendered unnecessary.

Therefore, we do not consider the use of implicit flow to be a downside, assuming standard safety measures -- such as the use of \texttt{POST} redirects with the ID Token -- are taken.\footnote{We also refer to previous work by Kroschewski and Lehmann~\cite{DBLP:journals/popets/KroschewskiL23} for additional discussion on the implicit flow as a privacy-enabling tool.}

\myparagraph{User obtains their pseudonym}
In traditional OpenID Connect, when using the authorization code flow, the ID Token (and the claims contained therein) are invisible to the user.
This means that the user does not learn the pseudonym that is exposed (in the \texttt{sub} claim) to the service provider.

When using OpenID Connect using implicit flow, as happens with BISON-augmented OIDC, the user learns the ID Token.
Since the user also knows $\R$, they can also derive their BISON pseudonym $\UID \cdot \AUDHASH$ using the same method as the SP.

We do not view this as a weakness of BISON;
  we believe that users \emph{in general} should have awareness of what information about them is exchanged.
Regardless, we still note this departure from the knowledge model of traditional OIDC.

However, we also note that the user still does not obtain their $\UID$;
 this information is always private to the IdP.

\myparagraph{Choice of cryptographic parameters}
When instantiating BISON-OIDC for our demonstrator, we use ristretto255 for simple reasons of operational convenience:
namely, ristretto255 had conveniently-available Java%
 \footnote{\url{https://github.com/cryptography-cafe/curve25519-elisabeth}}
and JavaScript%
 \footnote{\url{https://github.com/facebook/ristretto255-js}}
implementations.

In general, BISON can be instantiated over any prime-order group on which the DLP and DDHP are hard.
Prime-Order OPRFs have similar requirements;
  here, RFC 9497 suggests, in addition to ristretto255, the decaf448, P-256, P-384, and P-521 curve groups~\cite{DBLP:journals/rfc/rfc9497}.
For \enquote{real} practical adoption of BISON-OIDC, the NIST P-curves~\cite{nistcurves}, which are widely supported in cryptographic hardware, would likely be a more suitable choice.

The choice of SHA-512 also matches the CFRG draft, and is a natural fit for the ristretto255 group.
Since SHA-512 is a standardized cryptographic hash function, fast, secure implementations of it are widely available.

\section{Related Work}
\label{related-work}

\begin{table*}[t]
\caption{\label{tbl:relwork} Overview of privacy-focused SSO systems. Property supported \fully, partially supported \partially, or unsupported \unsupported}
\newcommand{\zerocite}[1]{\makebox[0pt][l]{\cite{#1}}}
\newcommand{\zerosuperscript}[1]{\makebox[0pt][l]{\textsuperscript{#1}}}
\normalsize
 \begin{center}

\begin{tabular}{ll|cc|cccc}
 &  & \makecell{Relationship\\Anonymity} & \makecell{Pseudonym\\Unlinkability} & Sybil Resistance & Stateless & \makecell{Client\\Unmodified} & \makecell{Basic\\Cryptography}\\
\hline
OIDC & \cite{OIDC} & \unsupported & \unsupported & \fully & \fully & \fully & \fully\\
OIDC w/ PPID & \cite[8.1]{OIDC} & \unsupported & \fully & \fully & \fully & \fully & \fully\\
Shibboleth & \cite{Shibboleth} & \unsupported & \fully & \fully & \fully & \fully & \fully\\
\hline
PseudoID & \cite{PseudoID10} & \partially & \fully & \unsupported & \unsupported & \unsupported & \partially\\
SPRESSO & \cite{DBLP:conf/ccs/FettKS15} & \fully & \unsupported & \fully & \fully & \fully & \fully\\
POIDC & \cite{DBLP:conf/ccs/HammannSB20} & \partially\zerosuperscript{1} & \partially\zerosuperscript{2} & \partially & \fully & \fully & \partially\zerosuperscript{2}\\
UPPRESSO & \cite{DBLP:journals/corr/abs-2110-10396} & \partially\zerosuperscript{1} & \fully & \fully & \fully & \fully & \fully\\
EL PASSO & \cite{DBLP:journals/popets/ZhangKSZR21} & \fully & \fully & \partially & \unsupported & \partially & \unsupported\\
AIF & \cite{DBLP:journals/popets/KroschewskiL23} & \partially\zerosuperscript{1} & \unsupported & \fully & \fully & \fully & \unsupported\\
ARPSSO & \cite{DBLP:conf/esorics/HeLWWJ24} & \partially\zerosuperscript{1} & \fully & \fully & \fully & \fully & \unsupported\\
OPPID & \cite{PETS:KroLehOez25} & \partially\zerosuperscript{1} & \fully & \fully & \fully & \fully & \unsupported\\
\hline
BISON &  & \fully & \fully & \fully & \fully & \partially\zerosuperscript{3} & \fully\\
\end{tabular}

\end{center}
{\raggedright 1. Only when assuming a very generous definition of a honest-but-curious identity provider. \par}
{\raggedright 2. Pairwise POIDC requires ZK-proofs, but offers unlinkability. \par}
{\raggedright 3. We discuss OIDC protocol awareness in browsers in \Cref{2:discussion}. \par}
\end{table*}

We survey existing protocols for delegated authentication, and in \Cref{tbl:relwork}, compare them against our privacy goals (cf. \Cref{1:goals}).
Further, we evaluate their compatibility;
  we check whether they require modifications to the browser, and whether they solely rely on basic cryptography.
We discuss these evaluation criteria further in \Cref{sec:appendix:evalcriteria}, and show our evaluation of each work in \Cref{sec:appendix:eval}.

We begin with protocols that are currently, or were previously, being used in real-world applications.

Baseline \emph{OpenID Connect} (OIDC)~\cite{OIDC} is not private.
It discloses the service provider's identity to the IdP, and the user's linkable identity to the SP.
OIDC optionally supports \emph{Pairwise Pseudonymous Identifiers} (PPIDs) to enable unlinkability between colluding SPs.\footnote{See also \Cref{2:background}.}
This requires the IdP to be aware of the SP's identity.

\emph{Shibboleth} is a SSO system based on SAML. It supports pseudonym unlinkability by means of \enquote{persistent identifiers}, which are unique to the combination of SP, IdP, and user~\cite{ShibbOIDCDocs}.
Just like OpenID Connect's PPIDs, the SP's identity must be known to the IdP.

\emph{BrowserID}~\cite{BrowserID} was a short-lived initiative by the Mozilla Foundation (branded as \textit{Mozilla Persona}) to embed secure delegated authentication in the browser.
While it succeeded in protecting User-SP relationship anonymity, it did not aim to provide pseudonym unlinkability, identifying users by their email address.
Further, Fett et al. discovered attacks against BrowserID~\cite{DBLP:journals/corr/FettKS14}.
They also exploit design flaws in BrowserID to allow arbitrary parties to query users' login status~\cite{DBLP:conf/esorics/FettKS15}.
Mozilla decommissioned the service in 2016.

Next, we will move on from protocols employed in existing software, and list approaches explored in academic literature.

\emph{PseudoID}~\cite{PseudoID10} uses blind signatures to extend OIDC's precursor OpenID with privacy features.
It introduces a \textit{blind signer} that issues pseudonym tokens to a user, that the user then uses to authenticate at the IdP.
This enables unlinkability since a user can use a separate pseudonym for each SP.
However, while the IdP does not learn which ``\textit{real}'' user is authenticating at a specific SP, it nevertheless can link multiple visits of the same \textit{pseudonymous} user.

BrowserID inspired \emph{SPRESSO}~\cite{DBLP:conf/ccs/FettKS15}.
It provides user-SP relationship anonymity, but not pseudonym unlinkability, as email addresses continue to be used as user identifiers.

\emph{POIDC}~\cite{DBLP:conf/ccs/HammannSB20} extends OIDC to prevent the IdP from learning at which SPs users authenticate.
It then extends this to \emph{pairwise POIDC}, which uses zero-knowledge proofs (ZKPs) to prevent colluding SPs from tracking users. %
Zero-knowledge proofs introduce nontrivial complexity to implementations.
This complexity, and a lack of standardization, often hinders a protocol's adoption in practice, or inclusion in other standards~\cite{DBLP:journals/iacr/TangSWCTS24}.
Further, ZKPs have non-trivial calculation overhead, exceeding that of BISON by orders of magnitude;
pairwise POIDC with SHA-256 requires 110ms, as opposed to BISON's 3ms (see \Cref{1:evaluation}).
This is needed, as in POIDC the IdP does not participate in the computation of the pseudonym.
Further, POIDC avoids browser modifications by requiring the SP and IdP to provide trustworthy JavaScript code to the user.
This strains the definition of a honest-but-curious identity provider;
  instead of merely performing authentication truthfully, POIDC's IdPs actively act contrary to their own data collection interests.
  By contrast, BISON does not expose sensitive information to IdP-controlled code.

The idea of using blinded evaluation to derive pseudonyms has also independently been explored by \emph{UPPRESSO}~\cite{DBLP:journals/corr/abs-2110-10396}.
However, its design is inflexible, and narrowly tailored to its use case.
By contrast, BISON is flexible, and can be easily adapted to fit any protocol.
UPPRESSO, like POIDC, asks the IdP to provide trustworthy JavaScript code to the user, which then acts to protect the user from the IdP's own behavior.
Further, in contrast to BISON, UPPRESSO assumes that $\AUDSTR$ is already a randomly-chosen group element.
Notably, if $\AUDSTR$ is not chosen randomly, this can introduce hidden backdoors in the form of known algebraic relationships between colluding SPs' pseudonyms.

\emph{EL PASSO}~\cite{DBLP:journals/popets/ZhangKSZR21} is a complex delegated authentication system that achieves both relationship anonymity and pseudonym unlinkability.
In exchange, it requires storing a user secret on the client device.
As discussed in \Cref{1:goals}, this is a significant downside in general; the same is true of EL PASSO.
If a user secret is lost, it envisions an out-of-band procedure to \enquote{replace} the user secret while updating a commitment at the IdP.
This limits Sybil resistance, as the user obtains a new set of pseudonyms as a result of the replacement.

The recent \emph{ARPSSO}~\cite{DBLP:conf/esorics/HeLWWJ24} achieves relationship anonymity by issuing an anonymous credential to the SP.
This credential is then used to perform a  two-party computation scheme to derive a (SP-specific) user identity.
While this computation relies on external JavaScript code, the authors note that this code can be provided by a third party. %
The proposal also supports the OIDC code-flow.

Another concurrent proposal is OPPID~\cite{PETS:KroLehOez25},
\textit{OPPID} is another concurrent proposal~\cite{PETS:KroLehOez25},
which achieves many of the same properties as BISON.
Like BISON, OPPID also cannot provide protection from colluding IdP and SPs.
In contrast to BISON, OPPID allows the IdP to authenticate the SP without relying on the user (device).
OPPID requires ZKP-friendly curves for its features.
In contrast, working on established standard curves allows BISON to be adopted in practice.
Implementers can use optimized, battle-tested libraries that they are already well-acquainted with.
Operations can be performed in existing cryptographic hardware.

Furthermore, due to employing ZKPs, OPPID's overhead of $\approx 26$ milliseconds also exceeds that of BISON by orders of magnitude.
In particular, most of these expensive computations are performed by the identity provider;
  this overhead is then multiplied by the number of parallel logins that large-scale IdP deployments see.

\emph{Other approaches} rely on blockchains~\cite{DBLP:journals/access/BelfaikSMSTS23},
distributed IdPs~\cite{DBLP:conf/ccs/AgrawalMMM18,DBLP:conf/eurosp/BaumFHLY20,DBLP:journals/popets/FrederiksenHPT23},
or hardware features like TEE or SGX~\cite{UPSSO,DBLP:conf/eurosp/XuYZF23}.
Reaching further, credential-based \enquote{user-centric} systems are often proposed an alternative to delegated authentication approaches, e.g.,
Idemix~\cite{DBLP:conf/eurocrypt/CamenischL01,DBLP:conf/ccs/CamenischH02},
U-Prove~\cite{DBLP:conf/securecomm/MostowskiV11},
ABC4Trust~\cite{DBLP:books/daglib/0039026},
and ZKlaims~\cite{DBLP:conf/icete/SchanzenbachKSB19}.
Those user-centric systems store the login information (credentials) on the user device and don't involve the IdP in the authentication process.
This need for a state on the user client, and the difficulty of key management and key backups~\cite{DBLP:conf/trustcom/HorandnerR19,DBLP:conf/iciss/HorandnerN19} have limited their practical adoption thus far~\cite{DBLP:conf/ccs/HammannSB20}.
This is also a limitation of other user-centric systems like Isaakidis' UnlimitID~\cite{DBLP:conf/wpes/IsaakidisHD16}.
Several works survey and evaluate privacy aspects of SSO systems in the wild, e.g.,~\cite{DBLP:conf/securecomm/WangBDD17,DBLP:journals/corr/abs-2302-01024,DBLP:journals/popets/DimovaGJ23}.
Finally, Brand{\~{a}}o et al.~\cite{DBLP:journals/popets/BrandaoCDa15} discuss nation-scale identification and authentication systems for credential management.

\ifnum\thepage>12

 \todoi{Page limit for non-Bibliography content is 12 pages! (currently \thepage)}

\fi

\clearpage

\bibliographystyle{ACM-Reference-Format} %
\bibliography{bib,dblp}

\appendix

\input{comparison.tex}

\input{impl_full.tex}
\subsection*{Availability}
We provide a full implementation of our OIDC extension, consisting of
BISON-aware service and identity providers (in Kotlin), and a browser extension for the Firefox browser. %
The source code is publicly available on GitHub\textsuperscript{\ref{fn:therepo}}, %
and a public demonstrator instance\textsuperscript{\ref{fn:thedemo}} is also available.
\ifanonymize
\else
\begin{anonsuppress}
\begin{acks}
  This work was supported by the European Union's research and innovation program 
  under grant agreements 
    \textnumero~101020416 (ERATOSTHENES) and 
    \textnumero~101168311 (LICORICE).
  We thank the anonymous reviewers for their valuable feedback throughout the review process.
  Further, we are grateful to Maria Eichlseder and Stefan Kreiner for inspirational discussions.
\end{acks}
\end{anonsuppress}
\fi

\par\noindent\rule{\columnwidth}{0.4pt}
 {
 \scriptsize
 \setlist{nosep}

 Paper history:
 \begin{itemize}
 \item 2023 October: Unsuccessful submission to USENIX '24 (Fall)
 \item 2024 May: Publication of revised paper as pre-print
 \item 2024 July: Unsuccessful Submission to NDSS '25
 \item 2024 July: Updated pre-print (DOI 10.48550/arXiv.2406.01518)
 \item 2024 September: Submission to AsiaCCS '25
 \item 2025 March: Submitted AsiaCCS '25 Major Revision
 \item 2025 April: Accepted at AsiaCCS '25
 \item 2025 July: Revised pre-print based on camera-ready version
 \item 2025 August: Presented at AsiaCCS '25
 \end{itemize}
}

\end{document}

%% file: comparison.tex
\newcommand{\aSystem}[2]{%
  \subsection[#1]{#1\if\relax\detokenize{#2}\relax\else~{\normalfont\small\cite{#2}, \citeyear{#2}}\fi}%
}

\newcommand{\cUnobservability}{Relationship Anonymity:}
\newcommand{\cUnlinkability}{Pseudonym Unlinkability:}

\newcommand{\cSybil}{Sybil Resistance:}
\newcommand{\cStateless}{Stateless:}
\newcommand{\cClientunmodified}{Client Unmodified:}
\newcommand{\cNoFancyCrypto}{Basic Cryptography:}

\newcommand{\eYes}{\fully{} }
\newcommand{\ePart}{\partially{} }
\newcommand{\eNo}{\unsupported{} }

\section{Evaluation Criteria}
\label{sec:appendix:evalcriteria}

The goal of this paper is to design a system that achieves the goals stated in \Cref{1:goals} (in addition to basic authentication-goals).
Correspondingly, we evaluate related work and our approach according to those goals, resulting in the following evaluation criteria:

\begin{description}
  \item[\cUnobservability] The identity provider (IdP) does not learn which service provider (SP) a user is authenticating with.
  \item[\cUnlinkability] Different SPs are not able to determine whether the same user's IdP account was used for authentication.
  \item[\cSybil] A single identity at the IdP corresponds to a single identity at each SP. A user cannot create multiple accounts at the same SP using a single IdP account.
  \item[\cStateless] The system does not require long-term storage on the user device. A user can authenticate using a fresh device without recovering cryptographic key material, or relying on any particular authentication factor.%
    \footnote{We find this to be a requirement in practice. Passwords are forgotten and have to be reset; security tokens are lost.}
  \item[\cClientunmodified] The system functions without modifications to the user's browser. While browser extensions are undesirable, they may be considered as a deployment trade-off to maximize compatibility.
    Some approaches avoid browser modifications by requiring the SP and IdP to provide trustworthy JavaScript code to the user.
    This strains the definition of a honest-but-curious identity provider, as these IdPs actively act contrary to their own data collection interests.
  \item[\cNoFancyCrypto] The system relies solely on widely deployed cryptographic primitives that are hardware-accelerated and supported by hardware security modules (HSMs). It avoids niche or experimental cryptographic constructions.
  \end{description}

\section{Detailed Evaluation}
\label{sec:appendix:eval}

We evaluate our approach and related work using the criteria discussed in \Cref{sec:appendix:evalcriteria}.
The results of this evaluation are also shown in \Cref{tbl:relwork} above.

\aSystem{OIDC}{OIDC}

\begin{description}
  \item[\cUnobservability] \eNo The IdP can see at which SP the user is authenticating.
  \item[\cUnlinkability] \eNo The SP receives an identifier which is the same for all SPs, thus allowing likability.
  \item[\cSybil] \eYes A unique account at the IdP results in a unique identifier shared with the SP.
  \item[\cStateless] \eYes The browser only redirects the user and does not store any state.
  \item[\cClientunmodified] \eYes OIDC only uses standard web browser features.
  \item[\cNoFancyCrypto] \eYes OIDC only uses basic signature schemes.
\end{description}

\aSystem{OIDC w/ PPID}{OIDC}

\begin{description}
  \item[\cUnobservability] \eNo The IdP can see at which SP the user is authenticating.
  \item[\cUnlinkability] \eYes Distinct SPs receive distinct identifiers for the same user. The Pairwise Identifier algorithm is not reversible by any party other than the IdP.
  \item[\cSybil] \eYes The Pairwise Identifier algorithm is deterministic.
  \item[\cStateless] \eYes The browser only redirects the user and does not store any state.
  \item[\cClientunmodified] \eYes OIDC only uses standard web browser features.
  \item[\cNoFancyCrypto] \eYes OIDC only uses basic signature and hash schemes.
\end{description}

\aSystem{Shibboleth}{Shibboleth}

\begin{description}
  \item[\cUnobservability] \eNo The IdP can see at which SP the user is authenticating.
  \item[\cUnlinkability] \eYes Shibboleth supports ``persistent identifiers'', which are unique for a SP, IdP, and user.
  \item[\cSybil] \eYes A unique account at the IdP results in a unique identifier shared with the SP.
  \item[\cStateless] \eYes The browser only redirects the user and does not store any state.
  \item[\cClientunmodified] \eYes Shibboleth uses only standard web browser features.
  \item[\cNoFancyCrypto] \eYes Shibboleth only uses basic signature and hash schemes.
\end{description}

\aSystem{PseudoID}{PseudoID10}

\begin{description}
  \item[\cUnobservability] \ePart The user authenticates at the IdP using a blinded access token retrieved from a token service (blind signer). As a consequence, the IdP sees the SP identifiers but does not learn which user is authenticating. However, the IdP learns the user's pseudonym in the process, allowing multi-show tracking.
  \item[\cUnlinkability] \eYes It is possible to use separate pseudo-\\nyms for each SP.
  \item[\cSybil] \eNo A user can create multiple pseudonyms and use them for the same SP.
  \item[\cStateless] \eNo Users retrieve blinded access tokens from the token service (blind signer), which they need to store and manage.
  \item[\cClientunmodified] \eNo Users need to manage their blinded access tokens locally, which is not supported by standard web browsers.
  \item[\cNoFancyCrypto] \ePart The approach relies on blind signatures, e.g., Chaum’s RSA blind signatures.
\end{description}

\aSystem{SPRESSO}{DBLP:conf/ccs/FettKS15}

\begin{description}
  \item[\cUnobservability] \eYes The IdP only receives the SP identifier in encrypted form.
  \item[\cUnlinkability] \eNo Email addresses are used as user identifiers, thus users are linkable among different SPs.
  \item[\cSybil] \eYes The user's identifier at the SP (i.e., email address) is directly used as user identifier.
  \item[\cStateless] \eYes The browser only redirects the user and does not store any state.
  \item[\cClientunmodified] \eYes SPRESSO \enquote{is based solely on standard HTML5 and web features and uses no browser extensions, plug-ins, or other executables}~\cite{DBLP:conf/ccs/FettKS15}.
  \item[\cNoFancyCrypto] \eYes SPRESSO only uses basic signature schemes.
\end{description}

\aSystem{POIDC}{DBLP:conf/ccs/HammannSB20}

The POIDC paper introduces two variant, POIDC and Pairwise POIDC, which we both evaluate here.

\begin{description}
  \item[\cUnobservability] \ePart \ePart Both POIDC variants blind the SP identifier before it is send to the IdP. However, in their paper the blinding happens in IdP-provided JavaScript code.
  \item[\cUnlinkability] \eNo \eYes POIDC includes a global (and thus linkable) user identifier in the ID token. Pairwise POIDC achieves unlinkability by also blinding the user's identifier on the client, and using a zero-knowledge proof to ensure correctness.%
  \item[\cSybil] \eYes \eNo POIDC exposes the global user identifier to the SP. Pairwise POIDC exposes the global identifier in blinded form and thus allows users to obtain multiple pseudonyms.
  \item[\cStateless] \eYes \eYes The browser only redirects the user and does not store any state.
  \item[\cClientunmodified] \eYes \eYes POIDC avoids browser modifications by performing client-side computations in server-provided JavaScript. This increases the trust requirements in the server and weakens the threat model.
  \item[\cNoFancyCrypto] \eYes \eNo Pairwise POIDC relies on zero-knowledge proofs. Using SHA-256 for blinding, POIDC's ZKBoo-based demo requires 55ms of time and 836 KB of space per blinding-step in the communication between user and IdP~\cite{DBLP:conf/ccs/HammannSB20}.
\end{description}

\aSystem{UPPRESSO}{DBLP:journals/corr/abs-2110-10396}

\begin{description}
  \item[\cUnobservability] \ePart Only an ephemeral SP identity is revealed to the IdP. However, in their paper the computation happens in SP- and IdP-provided JavaScript code.
  \item[\cUnlinkability] \eYes A SP-specific pseudo-identifier is derived from the user's global identifier. %
  \item[\cSybil] \eYes The calculation of the SP-specific pseudo-identifier is deterministic.
  \item[\cStateless] \eYes The browser only redirects the user and does not store any state.
  \item[\cClientunmodified] \eYes UPPRESSO avoids browser modifications by performing client-side computations in server-provided JavaScript.
  \item[\cNoFancyCrypto] \eYes UPPRESSO relies on signature schemes and ECC computations.
\end{description}

\aSystem{EL PASSO}{DBLP:journals/popets/ZhangKSZR21}

\begin{description}
  \item[\cUnobservability] \eYes By using anonymous credentials, the IdP is not involved in the authentication process at the SP. %
  \item[\cUnlinkability] \eYes Credentials are randomized by the user before presenting to the SP.
  \item[\cSybil] \ePart If users loose their stored secrets, they recover by retrieving a new pseudonym.
  \item[\cStateless] \eNo Requires storing a user secrets on the client device.
  \item[\cClientunmodified] \ePart It is deployed as WebAssembly client module by the IdP.
  \item[\cNoFancyCrypto] \eNo Utilizes anonymous (attribute-based) credentials and Pointcheval-Sanders signatures.
\end{description}

\aSystem{Authenticated Implicit Flow (AIF)}{DBLP:journals/popets/KroschewskiL23}

The paper introduces three variants:
Implicit Flow with standard signatures ($AIF_{SIG}$),
adapted POIDC ($AIF_{COM}$),
and their contribution $AIF_{ZKP}$.
Here we only evaluate $AIF_{ZKP}$.

\begin{description}
  \item[\cUnobservability] \ePart $AIF_{ZKP}$ hides the SP identifier and uses zero-knowledge proofs to ensure correctness. However, this computation either requires a browser modification or IdP-provided JavaScript code.
  \item[\cUnlinkability] \eNo The IdP generates a token that contains the user's global identifier. %
  \item[\cSybil] \eYes The identity token contains the user's global identifier.
  \item[\cStateless] \eYes AIF builds on OIDC's implicit flow. The browser only redirects the user and does not store any state.
  \item[\cClientunmodified] \eYes It avoids browser modifications by performing client-side computations in server-provided JavaScript.
  \item[\cNoFancyCrypto] \eNo $AIF_{ZKP}$ relies on zero-knowledge proofs to ensure correctness and enable SP accountability.
\end{description}

\aSystem{ARPSSO}{DBLP:conf/esorics/HeLWWJ24}

\begin{description}
  \item[\cUnobservability] \ePart SP authenticated using anonymous credentials. However, the computation relies on server-provided JavaScript code.
  \item[\cUnlinkability] \eYes Scoped pseudonym generated using the SP's credential and a  two-party secure computation scheme. The computation also supports the OIDC code flow. %
  \item[\cSybil] \eYes The pseudonym is deterministically computed using the user's identifier and the SP identifier.
  \item[\cStateless] \eYes The browser only redirects the user and does not store any state.
  \item[\cClientunmodified] \eYes It avoids browser modifications by performing client-side computations in server-provided JavaScript.
  \item[\cNoFancyCrypto] \eNo ARPSSO relies on Pointcheval-Sander signatures-based anonymous credentials issued to the SP.
\end{description}

\aSystem{OPPID}{PETS:KroLehOez25}

\begin{description}
  \item[\cUnobservability] \ePart The IdP only receives the blinded SP identifier. While not stated in the paper, we assume it follows its predecessor~\cite{DBLP:journals/popets/KroschewskiL23} and relies on server-provided JavaScript code.
  \item[\cUnlinkability] \eYes Pseudonyms are SP-specific by using HashDH-style pseudonyms. %
  \item[\cSybil] \eYes Pseudonyms are deterministically derived from the user's global identifier.
  \item[\cStateless] \eYes The browser only redirects the user and does not store any state.
  \item[\cClientunmodified] \eYes While not stated in the paper, we assume it follows its predecessor~\cite{DBLP:journals/popets/KroschewskiL23} and avoids browser modifications by performing client-side computations in server-provided JavaScript.
  \item[\cNoFancyCrypto] \eNo Relies on Schnorr zero-knowledge proofs to ensure correctness and enable SP accountability. The implementation relies on Pedersen commitments and Pointcheval-Sanders signatures (on curve BLS12-381).
\end{description}

\aSystem{BISON}{}

\begin{description}
  \item[\cUnobservability] \eYes Only an ephemeral SP identity is revealed to the IdP.
  \item[\cUnlinkability] \eYes  A SP-specific pseudonym is derived from the user's global identifier.
  \item[\cSybil] \eYes BISON identifiers are stable. Thus, a specific global user identifier always results in the same pseudonym at the same SP.
  \item[\cStateless] \eYes The browser only redirects the user and does not store any state.
  \item[\cClientunmodified] \ePart Identifier-transformations are performed on the client side, requiring either a browser modification or a browser extension.  BISON can optionally avoid browser modifications by performing client-side computations in server-provided JavaScript. This increases compatibility but weakens the threat model.
  \item[\cNoFancyCrypto] \eYes BISON only uses basic signature schemes and ECC computations on standardized curves.
\end{description}

%% file: impl_full.tex
\section{Demonstrator Implementation Details}
\label{sec:appendix:impldetails}

Our demonstrator implementation consists of three components: identity provider, service provider, and browser extension.
The two server components are written in Kotlin,\footnote{\url{https://kotlinlang.org/}} utilizing the Ktor server framework.\footnote{\url{https://ktor.io/}}
OpenID Connect functionality is provided by the Nimbus OAuth 2.0 SDK.\footnote{\url{https://connect2id.com/products/nimbus-oauth-openid-connect-sdk}}

Both servers have minimal business logic, which is self-contained in their respective \texttt{Main.kt} file.
They use FreeMarker\footnote{\url{https://ktor.io/docs/server-freemarker.html}} templates for pages.

\subsection{Identity Provider}
The Identity Provider generates an ephemeral ECDSA keypair over the P-256 curve on startup.
This keypair will be used to sign ID tokens.
It is published, alongside other parameters (including the \texttt{pairwise\_subject\_types} value \texttt{[\enquote{bison}]}), using OIDC Discovery~\cite{oidc-discovery}.

HTTP GET requests to \texttt{/login}, which is the OIDC authorization endpoint, are parsed.
The provided OIDC parameters are printed out, and a choice between identities is presented.
Once a choice has been made, this triggers a POST form submission to \texttt{/login}.
Based on the choice, a hardcoded constant $\UID$ is looked up.

If the associated OIDC parameters included \texttt{pairwise\_subject\_type=bison}, the IdP base64url-decodes the \texttt{client\_id} (which is $\BLINDINPUT := \BLIND(\AUDHASH)$).
It then calculates $\BLINDOUTPUT := \EVAL(\BLINDINPUT, \UID)$.
This result is base64url-encoded and used as the \enquote{subject} (\texttt{sub}) claim of the ID Token.
The base64url-encoded $\BLINDINPUT$ is also stored in the ID token as the \enquote{audience} (\texttt{aud}) claim.

If the OIDC parameters did not include the BISON marker parameter, \enquote{normal} pairwise pseudonym derivation is performed instead.
The \texttt{sub} claim is set to the SHA-512 hash of the concatenation of the \texttt{client\_id} and $\UID$.

In either case, the resulting ID token is then signed using the ephemeral P-256 ECDSA keypair that was generated on startup.
The result is inserted into a self-submitting HTML form as per the \texttt{form\_post} response mode.

\subsection{Service Provider}
The Service Provider requests the Identity Provider's OIDC Discovery metadata on startup.
This allows is to retrieve the trusted ECDSA public key.

By default, the service provider presents a page instructing the user to click a button to log in.
Clicking the button redirects to \texttt{/auth}.
The page handler for \texttt{/auth} then creates an OIDC authentication request.
\texttt{client\_id} is set to the current web origin, and a random \texttt{nonce} is both generated and stored locally.
The page then sends a HTTP 302 redirect to redirect the user to the identity provider.

Once the user returns from the identity provider, the resulting ID token is inspected.
If it contains a \texttt{pairwise\_subject\_type} claim with value \texttt{bison}, BISON validation is performed.
In either case, the signature, \texttt{aud} and \texttt{nonce} claims are all checked.
If the ID token passes validation, the value of \texttt{sub} is used to calculate the user's pseudonym.

In the case of \enquote{regular} OIDC, the value itself is already the pseudonym.
For BISON-OIDC, the value is base64url-decoded to obtain $\BLINDOUTPUT$;
  the pseudonym is then $\UNBLIND(\BLINDOUTPUT, \BLIND)$.

This pseudonym is then displayed to the user.

\subsection{Browser Extension}
BISON-OIDC requires a protocol-aware browser.
We implement a browser extension that enables this for the Firefox browser.
In a practical roll-out of BISON-OIDC, this functionality should be integrated into browsers directly.
We refer to \Cref{2:discussion} for further discussion of this.

Our browser extension requests the \texttt{webRequestBlocking} permission, which allows it to intercept and redirect web requests are they are being made.
We only request this permission for \path{http://localhost/}, as well as our demonstator domains.
This is purely because we do not wish to concern readers by requesting excessive permissions;
  we stress that the extension would work seamlessly on the wider internet by requesting appropriate wildcard permissions.

Once installed and running, the extension installs two separate handlers that intercept web requests.

The first intercepts requests to any domain, and checks whether they are OIDC authorization requests (\texttt{scope=openid}).
If they are, they are checked for BISON support (\texttt{pairwise\_subject\_types} includes \texttt{bison}).
If so, the audience ID is validated as a registrable superdomain of (or the same domain as) the current web origin.
The audience ID is either the special \texttt{audience\_id} parameter if present, which functions as an override, or otherwise the \texttt{client\_id}.
If any of these criteria are not met, the request proceeds as usual.
If all criteria are met, BISON blinding is performed.
The \texttt{nonce} parameter is also replaced with the SHA-512 hash of the concatenation of the current web origin and the server-specified nonce.
All other identifying parameters (such as \texttt{audience\_id} are stripped.
The BISON marker (\texttt{pairwise\_subject\_type=bison}) is added.
The \texttt{redirect\_uri} is replaced with \path{https://anonymous.invalid/bison}.
The request is then allowed to proceed with the modified parameters.

The second intercepts requests to \path{https://anonymous.invalid/bison};
  in other words, OIDC responses to requests blinded by the extension.
The original \texttt{redirect\_uri} value is looked up, the \texttt{blind} is added, and the request is allowed to proceed to the original target page.